\documentclass[journal]{IEEEtran}

\usepackage{amsmath,graphicx}
\usepackage{algorithm,algorithmic}
\usepackage{cite,amssymb,amsfonts,color,textcomp}
\usepackage{bm}
\usepackage{balance}
\usepackage{booktabs}
\usepackage{amsthm}
\usepackage{setspace}

\makeatletter
\newcommand*{\rom}[1]{\expandafter\@slowromancap\romannumeral #1@}
\makeatother

\def\gammabf{\boldsymbol \gamma }

\def\etabf{\boldsymbol \eta }

\def\phibf{\boldsymbol \phi }

\def\abf{{\bf a}}

\def\dbf{{\bf d}}

\def\fbf{{\bf f}}

\def\hbf{{\bf h}}

\def\qbf{{\bf q}}

\def\ubf{{\bf u}}
\def\vbf{{\bf v}}
\def\wbf{{\bf w}}
\def\xbf{{\bf x}}
\def\ybf{{\bf y}}
\def\zbf{{\bf z}}

\def\xbf{{\bf x}}
\def\ybf{{\bf y}}
\def\Abf{{\bf A}}

\def\Cbf{{\bf C}}

\def\Fbf{{\bf F}}

\def\Hbf{{\bf H}}
\def\Ibf{{\bf I}}

\def\Rbf{{\bf R}}

\def\Dc{{\cal D}}

\def\Fc{{\cal F}}
\def\Gc{{\cal G}}
\def\Hc{{\cal H}}

\def\Kc{{\cal K}}
\def\Lc{{\cal L}}

\def\Pc{{\cal P}}

\def\Sc{{\cal S}}

\def\Uc{{\cal U}}

\def\Zc{{\cal Z}}

\def\eg{{\it e.g.,\ \/}}

\def\ie{{\it i.e.,\ \/}}
\def\nn{\nonumber}

\def\tFbf{{\widetilde{\bf F}}}

\def\tq{{\tilde{q}}}
\def\tqbf{{\tilde{\bf q}}}
\def\te{{\tilde{e}}}

\def\Re{\mathfrak{R}\mathfrak{e}}
\def\Im{\mathfrak{I}\mathfrak{m}}

\newcommand{\sss}{\scriptscriptstyle}
\def\sssQ{{\sss \text{Q}}}

\theoremstyle{definition}

\newtheorem{remark}{Remark}

\newtheorem{proposition}{Proposition}
\newtheorem{definition}{Definition}

\newenvironment{mylist}%
{\begin{list}{}%
    {%
      \setlength{\itemindent}{-5pt}%
      \setlength{\leftmargin}{12pt}%
      \setlength{\parsep}{\parskip}
      \setlength{\labelsep}{5pt}
      \setlength{\itemsep}{2pt}}}%
  {\end{list}}

\begin{document}

\title{\huge Ultra-Low-Complexity   Algorithms with Structurally Optimal Multi-Group Multicast Beamforming in Large-Scale Systems}

\author{Chong~Zhang,~\IEEEmembership{Student~Member,~IEEE,}
        Min~Dong,~\IEEEmembership{Senior~Member,~IEEE,}
        and~Ben~Liang,~\IEEEmembership{Fellow,~IEEE}%
\thanks{Part of this work was presented at the IEEE International Conference on Acoustics, Speech and Signal Processing, 2021 \cite{Zhang&etal:2021ICASSP}.

Chong Zhang and Ben Liang are with the Department of Electrical and Computer Engineering, University of Toronto, Toronto,
ON M5S 1A1, Canada (e-mail: chongzhang@ece.utoronto.ca; liang@ece.utoronto.ca).

Min Dong is with the Department of Electrical, Computer and Software Engineering, Ontario Tech University, Oshawa,
ON L1G 0C5, Canada (e-mail: min.dong@ontariotechu.ca). } }%

\maketitle

\begin{abstract}
In this work, we propose ultra-low-complexity design solutions for multi-group multicast beamforming in large-scale systems.
For the quality-of-service (QoS) problem, by utilizing the optimal multicast beamforming structure obtained recently
in \cite{Dong&Wang:TSP2020},
we convert the original  problem into a non-convex weight optimization problem of a lower dimension and propose two fast  first-order algorithms to solve  it. Both algorithms are based on successive convex approximation (SCA) and provide fast iterative updates to solve each SCA subproblem. The first algorithm uses a  saddle point reformulation
in the dual domain and applies the extragradient method with an adaptive step-size procedure to find the saddle point
with simple closed-form updates.
 The second algorithm adopts the alternating direction method of multipliers (ADMM) method by converting each SCA subproblem into a favorable ADMM
structure. The structure leads to simple  closed-form ADMM updates, where the problem in each update block can be further  decomposed into parallel  subproblems of small sizes, for which closed-form solutions are obtained.
 We also propose efficient initialization methods to obtain favorable initial points that facilitate  fast convergence.
Furthermore,
taking advantage of the proposed fast algorithms, for the max-min fair (MMF) problem, we  propose
a simple closed-form scaling scheme that directly uses the solution obtained from the QoS problem, avoiding the  conventional computationally expensive method that iteratively solves the inverse QoS problem.
We further develop lower and upper bounds on the performance of this scaling scheme. Simulation results show that the proposed algorithms  offer near-optimal performance with substantially
lower computational complexity than the state-of-the-art algorithms for large-scale systems.
\end{abstract}


\section{Introduction}
\label{sec:intro}
Content distribution through wireless multicasting has been increasingly popular among new wireless services and applications
and is expected to dominate future wireless traffic.
Multi-antenna multicast beamforming is an efficient transmission technique to support high-speed content distribution to multiple user groups simultaneously.
 With   massive multiple-input multiple-output (MIMO)  being the essential technology  for future networks \cite{Larsson&Edfors&Tufvesson&Marzetta:ICM:14},
 it is critical for multicast beamforming solutions to be scalable to meet the ultra-low complexity requirement for  large-scale systems.

Downlink multicast beamforming has been studied for
traditional multi-antenna systems in various scenarios, such as
single-group or multi-group multicasting  \cite{Sidiropoulos&etal:TSP2006,Karipidis&etal:TSP2008,Ottersten&etal:TSP14},
 multi-cell networks \cite{Jordan&Gong&Ascheid:Globecom09,Xiang&Tao&Wang:IJWC:13},  relay networks \cite{Bornhorst&etal:2012TSP,DongLiang:CAMSAP13}, and cognitive radio networks \cite{Phan&etal:2009TSP,Huang&etal:2012TSP}.
Multicast beamforming problems are challenging to solve, as they are generally non-convex and NP-hard \cite{Sidiropoulos&etal:TSP2006}.
Semi-definite relaxation (SDR)  has been a prior state-of-the-art method considered in the existing works \cite{Luo&etal:2010SPM}.
For traditional multi-antenna systems with relatively small  problem sizes,
SDR provides  a good approximate solution
\cite{Sidiropoulos&etal:TSP2006,Karipidis&etal:TSP2008,Ottersten&etal:TSP14,Xiang&Tao&Wang:IJWC:13}.
However,
SDR is not a scalable method for  large-scale problems, where the computational complexity becomes very high
 and the performance deteriorates significantly as  the problem size grows.

With the increasing number of transmit antennas,  successive convex approximation (SCA)
\cite{Marks&Wright:OperReas1978} has become a more attractive
approach for solving the multi-group multicast beamforming problems due to its
 computational and performance advantages over SDR \cite{Tran&etal:SPL2014,Mehanna&etal:2015,Christopoulos&etal:SPAWC15}.
 SCA-based methods convexify the non-convex problem
 into a sequence of convex approximation subproblems, where the convex subproblems are typically solved by  the interior-point method (IPM) \cite{Boyd&Vandenberghe:book2004}. However, IPM is a second-order algorithm. For computing a multicast beamforming solution in large-scale massive MIMO systems,    using IPM still results in a relatively high computational
complexity.

Following this, several methods have been proposed    to further reduce the computational complexity
at each SCA iteration. For the single-group case, first-order methods have been proposed to solve the convex subproblems
\cite{Konar&Sidiropoulos:TSP2017,Ibrahim&etal:TSP2020}.
However, these methods  are not directly applicable to multiple groups with inter-group interference. For the multi-group scenario with per-antenna power constraints, the
alternating direction method of multipliers (ADMM) \cite{Chen&Tao:ITC2017}
has been considered for solving the subproblem in each SCA iteration.
In  \cite{Sadeghi&etal:TWC17}, zero-forcing pre-processing for interference elimination has been proposed to reduce the multi-group case to a single-group equivalent problem, for which SCA is applied. These methods provide much lower complexity than the original SCA-based method. However,
since the dimension of beamforming vectors is dictated by  the number of antennas, the computational complexity of these methods still  grows  with the number of antennas in polynomial time. This  renders these methods still  computationally costly for massive MIMO systems. Alternatively, for  multi-cell systems, low-complexity beamforming schemes using weighted maximum ratio transmission (MRT) in combination with SDR have  been developed, where only the weights, one for each user, need to be optimized, and thus the size of the optimization problem is reduced  \cite{Yu&Dong:ICASSP18,Yu&Dong:SPAWC18}.
Despite all the above advancements in computational algorithms, they do not optimally utilize the multicast beamforming structure.

The optimal multi-group multicast beamforming structure has been recently obtained in \cite{Dong&Wang:TSP2020}. It is shown that the optimal solution is a weighted minimum mean-square error (MMSE) filter with an inherent low-dimensional structure for the unknown weights  to be computed.
 With this structure, the multicast beamforming  problem can be transformed into a weight optimization problem of a much lower dimension,  independent of the  number of antennas.
As a result, the solution can be computed with significantly lower computational complexity, no longer growing with the number of    antennas \cite{Dong&Wang:TSP2020}.
Thus, it offers design opportunities for
computationally efficient algorithms for large-scale massive MIMO systems.
However, \cite{Dong&Wang:TSP2020} still adopted  the conventional  IPM-based SCA method for the weight optimization,
whose computational complexity still does not scale well with the total number of users.
 Our goal in this work is to develop scalable  first-order fast algorithms  for  large-scale systems that exploit both the optimal beamforming structure and optimization techniques.

Multicast beamforming  problems considered in all the above-mentioned works typically  are cast
into two problem formulations: a quality-of-service (QoS) problem for transmit power minimization
with  signal-to-interference-and-noise (SINR) guarantees for all users,
or a max-min fair (MMF) problem for maximizing the minimum SINR among all users subject to a transmit power budget.
Although both types of problems are non-convex and NP-hard,
the MMF problem is  more complicated to solve than the QoS problem.
Typically,
the solution to the  MMF problem is obtained through
 iteratively solving its inverse QoS problem along with a bi-section search over the value of minimum SINR   \cite{Karipidis&etal:TSP2008, Ottersten&etal:TSP14, Christopoulos&etal:SPAWC15, Chen&Tao:ITC2017, Dong&Wang:TSP2020}.
This additional layer of bi-section iterations results in
 high computational complexity for the  MMF problem in large-scale systems.
Therefore, developing a low-complexity method to obtain a good solution for the MMF problem is also critically important.

 Besides the  above-mentioned works  on algorithm development for multicast beamforming design,  asymptotic  multicast beamforming  in massive MIMO systems has  been analyzed  in \cite{Xiang&etal:2014JSAC,Sadeghi&Yuen:2015Globecom}
  without inter-group interference consideration,
  and in  \cite{Dong&Wang:TSP2020}   with  inter-group interference consideration.
Multicast beamforming has also been investigated for energy efficiency maximization \cite{Tervo:TSP18} and for joint unicast and multicast transmission in massive MIMO systems \cite{Sadghi&etal:IEEE_J_WCOM18,Mohammadietal:SPAWC21}.  For overloaded systems with fewer antennas than the users,
 the rate-splitting-based multicast beamforming strategies have been
proposed  \cite{Joudeh&Clerckx:2017TWC,Tervo&etal:2018SPAWC,Chen&etal:2020TVT}.
These studies address different problems from the one considered in our work.

\subsection{Contributions}

In this paper, for downlink multi-group multicast beamforming  in large-scale massive MIMO systems, we propose two  fast first-order   algorithms  for the QoS problem, which are scalable in both antenna and user dimensions.  Utilizing the optimal multicast beamforming structure, we convert the original QoS problem into a non-convex weight optimization problem of a much lower dimension. We then develop two fast algorithms to solve this weight optimization problem  to obtain our multicast beamforming solution.  The two algorithms are SCA-based methods, referred to as the extragradient-based SCA (ESCA) and ADMM-based SCA (ASCA).
They are developed using two different first-order optimization techniques. Using these algorithms, we also propose
a simple closed-form scaling scheme for  solving the MMF problem. The main  novelty and contribution of this work are summarized below:
\begin{mylist}
\item We  propose  ESCA to solve each SCA subproblem in the dual domain. In particular, we construct a saddle point reformulation of the dual problem, which can be further cast as a variational inequality problem for us to  apply the extragradient method \cite{Korpelevich:Matecon1976} to find the saddle point. Instead of directly considering the  primal problem,  our approach explores the dual problem where the extragradient method  can be used efficiently, and we obtain simple closed-form gradient updates in each updating step, which is the key for our  algorithm to  compute the solution with low complexity.
Furthermore, to avoid finding the Lipschitz constant of the gradient function to set the step size, which  is generally difficult to obtain, we adopt  a prediction-correction
procedure   for an adaptive step size in each update to ensure convergence to the saddle point for the optimal solution.
We also consider two initialization methods, a   fast   extragradient-based initialization method and an SDR-based method,  for generating favorable initial feasible points for ESCA to facilitate fast convergence.

\item We also propose  a fast ADMM-based algorithm for the SCA subproblems, referred to as ASCA. We transform each SCA subproblem into a favorable form to construct the  ADMM  procedure with two ADMM updating blocks. In particular, the structure of the transformed problem leads to simple  updates in each ADMM update block, where the problem in each update block can be further  decomposed into parallel  subproblems of small sizes, each of which yields a closed-form solution.
Thus, ASCA takes advantage of the closed-form updates in the ADMM procedure  for fast computation and is guaranteed to converge to the
optimal solution. The similar procedure is used to provide a  ADMM-based fast initialization method to facilitate  fast convergence of the algorithm.

\item Taking advantage of the  proposed fast algorithms for the QoS problems, we propose a simple closed-form scaling scheme to obtain a multicast beamforming solution for the MMF problem. The scheme directly scales the beamforming solution obtained from the QoS problem
to meet the transmit power budget.
It thus has substantially lower computational complexity  than the
conventional  method of iteratively solving the inverse QoS problem with bi-section search.
We also provide lower and upper bounds on the performance of the scaling scheme.

\item  Simulation results demonstrate
that both  ESCA and ASCA   with their  initialization methods provide near-optimal performance for the QoS problem.  Their computational complexity is substantially lower  than the state-of-the-art algorithms  for large-scale systems as  the number of antennas and users increases, demonstrating the algorithm scalability  in large-scale systems. Between the two algorithms, ASCA is preferable with faster computation for small to moderately large systems where the number of antennas is less than 300 and the number of users per group is less than 20, while ESCA provides faster computation for  larger systems.
In addition,  the proposed scaling scheme for the MMF problem is shown to  result in only a mild loss  to the optimal performance as the number of antennas becomes large but substantially faster to compute the solution.
\end{mylist}

We note that a multicast beamforming  problem under per-antenna power constraints has been considered in   \cite{Chen&Tao:ITC2017}, where an ADMM-based algorithm has been proposed. Besides the problem being different from ours, the optimal beamforming structure is not considered there, and the iterative algorithm targets directly computing the beamforming vector. The resulting algorithm in  \cite{Chen&Tao:ITC2017} has three ADMM updating blocks, while our ASCA contains two ADMM blocks.    In general, different from the existing algorithms \cite{Konar&Sidiropoulos:TSP2017,Ibrahim&etal:TSP2020,Chen&Tao:ITC2017,Sadeghi&etal:TWC17,Yu&Dong:ICASSP18,Yu&Dong:SPAWC18}, our proposed two algorithms, ESCA and ASCA, exploit both optimal multicast beamforming structure and the numerical optimization techniques, which lead to the simple closed-form updating steps to compute the multicast beamforming solution with  ultra-low computational complexity  for  large-scale systems.

\subsection{Organization and Notations}
The rest of this paper is organized as follows.
Section~\ref{sec:system_prob} presents the system model and problem formulation
for multi-group multicast beamforming.
Section~\ref{sec:sca_optimal} reviews the optimal multicast beamforming structure and the SCA method.
In Sections~\ref{sec:ESCA} and V, we present two fast algorithms, ESCA and ASCA, for the QoS problem.
In Section~\ref{sec:duality}, we propose a simple closed-form scaling scheme to find the solution for the  MMF problem.
Simulation results are provided in Section~\ref{sec:simulations},
and the conclusion is presented in Section~\ref{sec:conclusion}.

\textit{Notations}:
Hermitian, transpose, and conjugate are denoted as
$(\cdot)^{H}$, $(\cdot)^{T}$, and $(\cdot)^{*}$, respectively.
The real and imaginary parts
of a complex number are denoted as $\Re{\{\cdot\}}$ and $\Im{\{\cdot\}}$, respectively.
The Euclidean norm of a vector
is denoted as $\|\cdot\|$.
The notation $\xbf\succcurlyeq {\bf 0}$ indicates
element-wise non-negative. The identity matrix is denoted
as $\Ibf$.
The notation
$\zbf \sim \mathcal{CN}({\bf{0}},{\bf{C}})$ means $\zbf$ is
a complex Gaussian random vector with zero mean and covariance $\Cbf$.
 The non-negative real Euclidean space is denoted as $\mathbb{R}_{+}$.

\allowdisplaybreaks
\section{System Model and Problem Formulation}
\label{sec:system_prob}
We consider a downlink multi-group multicast beamforming scenario,
where the base station (BS) equipped with $N$ antennas provides service for $G$ user groups.
Each group receives a common message that is independent of the messages to other groups.
Denote the set of group indices by $\Gc \triangleq \{1,\cdots,G\}$.
Assume that there are $K_i$ single-antenna users in group $i$,
and the set of user indices in the group is denoted by $\mathcal{K}_{i} \triangleq \{1,\cdots,K_i\}, \, i\in\Gc$.
Define the total number of users in all groups as $K_{\text{tot}}\triangleq \sum_{i=1}^{G}K_i$.

Let $\hbf_{ik}\in \mathbb{C}^{N\times 1}$ be the channel vector from the BS to user $k$ in group $i$,
for $k\in \Kc_i$, $i\in \Gc$.
Let $\wbf_i\in \mathbb{C}^{N\times 1}$ be the multicast beamforming vector for group $i\in\Gc$.
The received signal at user $k$ in group $i$ is given by
\begin{align}
y_{ik}={\bf{w}}^{H}_{i}{\bf{h}}_{ik}s_{i}+\sum_{j\neq i}{\bf{w}}^{H}_{j}{\bf{h}}_{ik}s_{j}+n_{ik} \nonumber
\end{align}
where $s_i$ is the data symbol transmitted to group $i$ with $E\left(|s_{i}|^{2}\right) = 1$,
and $n_{ik}$ is the receiver additive white Gaussian noise with zero mean and variance $\sigma^{2}$.
The received SINR at user $k$ in group $i$ is expressed as
\begin{align}
\text{SINR}_{ik}=\frac{|\wbf_i^{H}\hbf_{ik}|^{2}}{\sum_{j\neq i}|{\bf{w}}_{j}^{H}\hbf_{ik}|^{2}+\sigma^2}.\nn
\end{align}
The transmit power at the BS is given by $P_{\rm t}$ $\triangleq\sum_{i=1}^{G} \|\wbf_i\|^2$.

Two types of problem formulations are typically considered for the multicast beamforming design.
The QoS problem is to minimize the transmit power while meeting the received SINR target
at each user, which is formulated as
\begin{align}
\Pc_o: & \min_{{\bf{w}}} \,\, \sum_{i=1}^{G} \|\wbf_i\|^2  \quad \text{s.t.} \, \ \text{SINR}_{ik}\geq \gamma_{ik},\,\, k\in \mathcal{K}_i, i\in \Gc
\label{SINR_constraint}
\end{align}
where ${\bf{w}}\triangleq[{\bf{w}}_{1}^H, \cdots, {\bf{w}}_{G}^H]^H$,
and $\gamma_{ik}$ is the SINR target for user $k$ in group $i$.
The other problem formulation is the (weighted) MMF problem, which is to maximize the minimum (weighted) SINR among all users
subject to the transmit power constraint at the BS:
\begin{align}
\Sc_o:  \max_{{\bf{w}}} \min_{i,k} &\,\, \frac{\text{SINR}_{ik}}{\gamma_{ik}}
&\text{s.t.}  \quad \sum_{i=1}^{G} \|\wbf_i\|^2\leq P
\label{MMF_constraint}
\end{align}
where $P$ is the transmit power budget at the BS, and $\gamma_{ik}>0$, $\forall i,k$, here serves as the weight to control the grade of service or fairness among users.
It is well-known that both $\Pc_o$ and $\Sc_o$ are non-convex and NP-hard.
The existing works have proposed various computational  optimization methods
to find good suboptimal solutions.
We will first focus on the QoS problem $\Pc_{o}$
and develop two fast first-order
algorithms to obtain the  solutions for $\Pc_{o}$.
Then, we discuss how to use the proposed algorithms to solve the MMF problem efficiently.

\section{Optimal Multicast Beamforming Structure and The SCA Method}
\label{sec:sca_optimal}

The optimal multicast beamforming structure has been
obtained recently in \cite{Dong&Wang:TSP2020}.
Under this structure, problem $\Pc_o$ is transformed into an equivalent
weight optimization problem of a much lower dimension that is independent of the number of antennas.
This leads to a significant computational saving and provides opportunities for efficient
algorithm designs for massive MIMO systems.
To facilitate our algorithm development later, we briefly describe the optimal multicast beamforming structure obtained in \cite{Dong&Wang:TSP2020},
the transformed weight optimization problem, and the SCA method for  the problem.

\vspace*{-.5em}
\subsection{Optimal Multicast Beamforming Structure}

It is shown in  \cite{Dong&Wang:TSP2020}
that the optimal solution to $\Pc_{o}$ is a weighted MMSE filter given by
\begin{align}
\wbf^{o}_{i} = {\bf{R}}^{-1}(\bm{\lambda}^{o})\Hbf_{i}\abf^{o}_{i},\,\, i\in \Gc
\label{optimal_QoS_structure}
\end{align}
where $\Hbf_{i} \triangleq \left[\hbf_{i1}, \cdots, \hbf_{iK_{i}}\right] \in \mathbb{C}^{N\times K_i}$ is the channel matrix for group $i$,
$\abf^{o}_{i}\in \mathbb{C}^{K_i\times 1}$ is the optimal weight vector for group $i$,
and
${\bf{R}}(\bm{\lambda}^o) \triangleq {\bf{I}}+\sum_{i=1}^{G}\sum_{k=1}^{K_{i}}\lambda_{ik}^o\gamma_{ik}\hbf_{ik}\hbf_{ik}^{H}\in \mathbb{C}^{N\times N}$ is the (normalized) noise plus weighted channel covariance matrix,
with $\bm{\lambda}^o\in \mathbb{R}^{K_{\text{tot}}\times 1}$ containing the optimal Lagrangian multipliers $\{\lambda_{ik}^o\}$ associated with the SINR constraints in \eqref{SINR_constraint}.
The $k$th weight element $a_{ik}^{o}$ in $\abf^{o}_{i}$ indicates
the relative significance of user $k$'s channel $\hbf_{ik}$ in the overall
group-channel direction $\Hbf_{i}\abf^{o}_{i}$.

To determine $\wbf^{o}$ in \eqref{optimal_QoS_structure},
we need to numerically compute both $\bm{\lambda}^{o}$ and $\{\abf^{o}_{i}\}$.
The parameter $\bm{\lambda}^{o}$ can be
approximately computed by the simple fixed-point iterative method proposed in \cite{Dong&Wang:TSP2020}.
Given $\bm{\lambda}$,
based on the optimal solution structure $\wbf^{o}_{i}$ in (\ref{optimal_QoS_structure}),
the original problem $\mathcal{P}_{o}$ is transformed
into the following equivalent weight optimization problem for $\{\abf_{i}\}$
\begin{align}
\Pc_1:  & \min_{{\abf}} \ \ \sum_{i=1}^{G} \|\Cbf_{i}\abf_{i}\|^2  \nonumber\\
  &  \text{s.t.} \  \ \frac{|\abf_{i}^{H}\fbf_{iik}|^{2}}{\sum_{j\neq i}|\abf_{j}^{H}\fbf_{jik}|^{2}+\sigma^2}\geq \gamma_{ik}, \, k\in \mathcal{K}_i, i\in \Gc
\label{Prb:P_2_constraint}
\end{align}
where $\abf \triangleq [\abf_{1}^H, \cdots, \abf_{G}^H]^H$,
 $\Cbf_{i} \triangleq \Rbf^{-1}(\bm{\lambda})\Hbf_{i}$ $\in \mathbb{C}^{N\times K_i}$,
 $\fbf_{jik} \triangleq \Cbf^{H}_{j}\hbf_{ik}$ $\in \mathbb{C}^{K_j\times 1}$, $k\in\Kc_{i}, i,j\in\Gc$.
Different from the original problem $\Pc_o$ with $GN$ variables,
the converted problem $\Pc_1$ has $K_{\text{tot}}$ variables,
which does not depend on the number of antennas $N$.
The problem dimension of $\Pc_{1}$ is much smaller  than that of $\Pc_{o}$
in massive MIMO systems with $K_i \ll N$.
The inherent  structure of the optimal multicast beamforming
solution in (\ref{optimal_QoS_structure})
paves the way for developing low-complexity fast algorithms for
multicast beamforming design in large-scale massive MIMO systems.

\vspace*{-.6em}

\subsection{Obtaining Weights  $\{\abf_i\}$ via SCA}
\label{subsec:SCA_prelimi}
The weight optimization problem
$\Pc_1$ is still a  non-convex NP-hard problem.
The SCA method can be adopted to solve $\mathcal{P}_{\rm 1}$,
which is guaranteed to converge to the local minimum \cite{Marks&Wright:OperReas1978}.
Specifically, given any auxiliary vector ${\bf{v}}_{i}\in \mathbb{C}^{K_i\times 1}$, $i\in\Gc$,
based on the inequality $(\abf_{i} - {\bf{v}}_{i})^{H}\fbf_{iik}\fbf_{iik}^{H}(\abf_{i} - {\bf{v}}_{i}) \ge 0$,
we have $|\abf_{i}^{H}\fbf_{iik}|^{2} \ge 2\Re{\{\abf_{i}^{H}\fbf_{iik}\fbf_{iik}^{H}{\bf{v}}_{i}\}}-|{\bf{v}}_{i}^{H}\fbf_{iik}|^{2}$.
Replacing the numerator of the SINR expression in \eqref{Prb:P_2_constraint}
by the right-hand side (RHS) of the above inequality,
we convexify the SINR constraint and change $\mathcal{P}_{\rm 1}$ to the following convex optimization problem
\vspace*{-.5em}
\begin{align}
 \Pc_{1\text{SCA}}({\bf{v}}): &
 \min_{\abf}\,\, \sum_{i=1}^{G} \|\Cbf_{i}\abf_{i}\|^2  \nonumber\\
             & \,\, \text{s.t.}  \,\,  \gamma_{ik}\sum_{j\neq i}|\abf_{j}^{H}\fbf_{jik}|^{2}+|{\bf{v}}_{i}^{H}\fbf_{iik}|^{2}+\gamma_{ik}\sigma^2  \nonumber\\
               &\qquad  -2\Re{\{\abf_{i}^{H}\fbf_{iik}\fbf_{iik}^{H}{\bf{v}}_{i}\}} \leq 0,\,\, k\in \mathcal{K}_i, i\in \Gc \nn
\end{align}
where ${\bf{v}} \triangleq [{\bf{v}}_{1}^H, \cdots, {\bf{v}}_{G}^H]^H$.
Note that the solution to $\Pc_{1\text{SCA}}(\vbf)$ is always feasible to $\Pc_{1}$.
By updating $\vbf$ with the solution $\abf$ to $\mathcal{P}_{1\text{SCA}}(\vbf)$,
we iteratively solve $\Pc_{1\text{SCA}}(\vbf)$ until convergence.

Since $\Pc_{1\text{SCA}}(\vbf)$ is convex,
it can be solved by IPM  available through
standard convex solvers.
However, IPM is a second-order algorithm (\ie based on the Hessian matrix of the objective function),
whose best computational complexity is  $O(K_{\text{tot}}^{3.5})$ and  worst is  $O(K_{\text{tot}}^{4})$.
Thus, iteratively solving  $\Pc_{1\text{SCA}}(\vbf)$
via IPM still incurs relatively high computational complexity for large-scale systems
 when $K_{i}$ is large.
To address this, for the rest of this paper,
 we develop two fast first-order algorithms to solve $\Pc_{1\text{SCA}}(\vbf)$ that
maintain a low complexity in computing the solution for large-scale systems.

\section{Extragradient-Based SCA Algorithm}
\label{sec:ESCA}
\subsection{Dual Saddle Point Reformulation}
\label{subsec:dual_saddle}

For the purpose of computation,
we rewrite problem $\Pc_{1\text{SCA}}({\bf{v}})$ using the real and imaginary parts of each complex quantity.
Define $\xbf_{i}\triangleq [\Re{\{\abf_{i}\}}^T, \Im{\{\abf_{i}\}}^T]^T$, $\ybf_{i}\triangleq [\Re{\{{\bf{v}}_{i}\}}^T, \Im{\{{\bf{v}}_{i}\}}^T]^T$.
Also, define
\begin{align}
\Abf_{i}\triangleq
\begin{bmatrix}
\Re{\{\Cbf_{i}\}} & \!\!\!\!\! -\Im{\{\Cbf_{i}\}} \\
\Im{\{\Cbf_{i}\}} & \!\ \Re{\{\Cbf_{i}\}}
\end{bmatrix}
,\tFbf_{jik}\triangleq
\begin{bmatrix}
\Re{\{\fbf_{jik}\}} & \!\!\!\!\! -\Im{\{\fbf_{jik}\}} \\
\Im{\{\fbf_{jik}\}} & \!\ \Re{\{\fbf_{jik}\}}
\end{bmatrix} \nn
\end{align}
for $k\in\Kc_{i}, i,j\in\Gc$.
Then, we have
$\|\Cbf_{i}\abf_{i}\|^{2} = \|\Abf_{i}\xbf_{i}\|^2$
and
$|\abf_{j}^{H}\fbf_{jik}|^{2}=\|\xbf_{j}^T\tFbf_{jik}\|^{2}=\xbf_{j}^{T}\Fbf_{jik}\xbf_{j}$,
where $\Fbf_{jik}\triangleq\tFbf_{jik}\tFbf_{jik}^{T}$, for $k\in \Kc_i$ and $j,i \in \Gc$.
Define $\xbf \triangleq [\xbf_{1}^T, \cdots, \xbf_{G}^T]^T$
and $\ybf \triangleq [\ybf_{1}^T, \cdots, \ybf_{G}^T]^T$.
With these new variables, $\Pc_{1\text{SCA}}({\bf{v}})$ can be equivalently expressed as
\begin{align}
&\Pc_{1\text{SCA}}^{\text{r}}(\ybf): \ \min_{\xbf}\,\, \sum_{i=1}^{G} \|\Abf_{i}\xbf_{i}\|^2 \nn \\
&\text{s.t.} \;\; \ybf_{i}^{T}\Fbf_{iik}\ybf_{i}+\!\gamma_{ik}\!\sum_{j\neq i}\xbf_{j}^{T}\Fbf_{jik}\xbf_{j}-2 \ybf_{i}^{T}\Fbf_{iik}\xbf_{i} +\gamma_{ik}\sigma^2\leq 0,\nonumber\\[-1em]
&\hspace*{16.5em} \,\, k\in\Kc_{i}, i\in \Gc. \label{2SCA}
\end{align}

We first describe the class of variational inequality problems \cite{Facchinei&Pang:book2003} below.

\begin{definition}[Variational Inequality]\label{def:VI}
Given $\Zc \subseteq \mathbb{R}^n$ and a mapping $\psi:\Zc \to \mathbb{R}^n$, the variational inequality is to find $\zbf \in \Zc$ satisfying
$\psi(\zbf)^T(\zbf'-\zbf)\ge 0$, $\forall~\zbf' \in \Zc$.
Operator $\psi(\cdot)$ is said to be monotone on $\Zc$ if $[\psi(\zbf)-\psi(\zbf')]^T(\zbf-\zbf') \ge 0$, $\forall~ \zbf,\zbf' \in  \Zc$.
The problem is  monotone if  operator $\psi(\cdot)$ is monotone.
\end{definition}

The projection methods belong to a class of iterative algorithms that solve the monotone variational inequality problems \cite{Facchinei&Pang:book2003}.
At each iteration, a projection method uses the updating step to compute the point
(\ie the value of the optimization variable)
and then projects it onto the feasible set of the problem to ensure the updated point is feasible.
Note that the projection method may not  be an efficient method. In general, the projection methods are only computationally cheap when the projection is easy to compute.

Note that  $\Pc_{1\text{SCA}}^{\text{r}}(\ybf)$ is convex, and the objective function is differentiable.
Let  operator $\psi(\xbf)$ be the gradient of the objective function of $\Pc_{1\text{SCA}}^{\text{r}}(\ybf)$.
Following the optimality criterion for a convex optimization problem,
$\psi(\xbf)$ is monotone, and thus $\Pc_{1\text{SCA}}^{\text{r}}(\ybf)$ is a monotone variational inequality problem.
However, it is difficult to find a closed-form expression for the projection onto the feasible set of $\Pc_{1\text{SCA}}^{\text{r}}(\ybf)$.
Thus, directly applying the projection method to solve $\Pc_{1\text{SCA}}^{\text{r}}(\ybf)$ is not computationally attractive.
To overcome this difficulty, we resort to the Lagrange dual problem of $\Pc_{1\text{SCA}}^{\text{r}}(\ybf)$.

Define
$\phi_{ik}(\xbf)\triangleq \ybf_{i}^{T}\Fbf_{iik}\ybf_{i}
+\gamma_{ik}\sum_{j\neq i}\xbf_{j}^{T}\Fbf_{jik}\xbf_{j}-2 \ybf_{i}^{T}\Fbf_{iik}\xbf_{i}  +\gamma_{ik}\sigma^2$ and  $\phibf_{i}(\xbf)\triangleq  \left[\phi_{i1}(\xbf),  \ldots, \phi_{iK_{i}}(\xbf)\right]^T$, for $k\in \Kc_{i}, i\in \Gc$.
The Lagrangian of $\Pc_{1\text{SCA}}^{\text{r}}(\ybf)$ is given by
\begin{align}\label{Lagrangian}
\Lc(\xbf, \bm{\eta})= \sum_{i=1}^{G}\left(\|\Abf_{i}\xbf_{i}\|^2+\bm{\eta}^{T}_{i}\phibf_{i}(\xbf)\right)
\end{align}
where $\eta_{ik}\ge 0$ is the dual variable associated with constraint \eqref{2SCA}, and   $\bm{\eta}\triangleq\left[\bm{\eta}_{1}^T, \cdots, \bm{\eta}_{G}^T\right]^T$
with $\bm{\eta}_{i}\triangleq\left[\eta_{i1}, \cdots, \eta_{iK_{i}}\right]^T$.
The Lagrange dual problem of $\Pc_{1\text{SCA}}^{\text{r}}(\ybf)$ is given by
\begin{align}
\Dc_{1\text{SCA}}^{\text{r}}(\ybf): \; \max_{\bm{\eta}\succcurlyeq {\bf 0}}\min_{\xbf} \sum_{i=1}^{G}\left(\|\Abf_{i}\xbf_{i}\|^2+\bm{\eta}^{T}_{i}\phibf_{i}(\xbf)\right).\nonumber
\end{align}

Let $\xbf^{o}$  and $\bm{\eta}^{o}$ be the primal
and dual optimal solutions for $\Pc_{1\text{SCA}}^{\text{r}}(\ybf)$.
 Since $\Pc_{1\text{SCA}}^{\text{r}}(\ybf)$ is convex and Slater's condition holds,
the strong duality holds. It follows that $\ubf^{o}\triangleq(\xbf^{o}, \bm{\eta}^{o})$
is a saddle point of the Lagrangian $\Lc(\xbf, \bm{\eta})$ \cite{Boyd&Vandenberghe:book2004}.
Define operator $g(\ubf)$ as
\vspace*{-.5em}
\begin{align}
g(\ubf)=g(\xbf, \bm{\eta})\triangleq\left[
\begin{matrix}
\nabla_{\xbf}\Lc(\xbf, \bm{\eta}) \\
-\nabla_{\bm{\eta}}\Lc(\xbf, \bm{\eta})
\end{matrix}
\right],\quad \ubf\in\Uc \label{operator}
\end{align}
where $\Uc \triangleq \mathbb{R}^{2K_{\text{tot}}}\times \mathbb{R}_{+}^{K_{\text{tot}}}$
is a  closed convex set.
Then, $\Dc_{1\text{SCA}}^{\text{r}}(\ybf)$ can be interpreted as finding the saddle point $\ubf^{o}$.
It is shown in \cite{Facchinei&Pang:book2003} that the problem of finding the saddle point $\ubf^{o}$ can be  cast as
the  variational inequality problem:  Find $\ubf^{o}\in\Uc$ that satisfies
the following variational inequality %
\begin{align}
g(\ubf^{o})^T(\ubf-\ubf^{o})\geq 0,\quad \forall~\ubf\in\Uc.
\label{var_ineq}
\end{align}

\subsection{Extragradient-Based SCA}
\label{subsec:ESCA_B}
To solve  problem \eqref{var_ineq},
one may consider the basic projection algorithm (BPA)  \cite{Facchinei&Pang:book2003},
which iteratively updates $\ubf$ using  $g(\ubf)$ and then projects it onto  $\Uc$.
However, the  convergence of BPA
requires operator $g(\ubf)$ to be strongly monotone  \cite{Facchinei&Pang:book2003}.
Following Definition~\ref{def:VI}, operator $\psi(\cdot)$ is said to be strongly monotone on $\Zc$, if there exists a constant $c>0$ such that  $[\psi(\zbf)-\psi(\zbf')]^T(\zbf-\zbf')$ $\ge$ $c\|\zbf-\zbf'\|^{2}$, $\forall~ \zbf,\zbf' \in \Zc$.
Since $\Lc(\xbf, \bm{\eta})$  is linear with respect to (w.r.t.) $\bm{\eta}$, operator
 $g(\ubf)$ is not strongly monotone on \,$\Uc$.
Thus, we cannot apply BPA to our problem due to no convergence guarantee.

Instead of BPA, in this work, we adopt  the extragradient method,  a variant of BPA,
 proposed by  Korpelevich in \cite{Korpelevich:Matecon1976}.
 Compared with  BPA,
the extragradient method can guarantee convergence for a monotone operator,
at the cost of an extra update-and-projection step at each iteration.
For   operator $g(\ubf)$ in \eqref{operator},
 since $\Lc(\xbf, \bm{\eta})$ is convex in $\xbf\in \mathbb{R}^{2K_\text{tot}}$ and  $-\Lc(\xbf, \bm{\eta})$ is convex in $\bm{\eta}\in \mathbb{R}_{+}^{K_\text{tot}}$,    $g(\ubf)$ is monotone on \,$\Uc$ by Definition~\ref{def:VI}.
Thus, we develop a fast iterative algorithm to solve $\mathcal{P}_{1\text{SCA}}^{\text{r}}(\ybf)$ by applying the extragradient algorithm  in the problem dual domain.

The updating procedure of the extragradient algorithm for finding the saddle point is summarized as follows \cite{Korpelevich:Matecon1976,Facchinei&Pang:book2003}: At iteration $n+1$,
\begin{align}
\bar{\xbf}^{(n)} &=  \xbf^{(n)}-\alpha\nabla_{\xbf^{(n)}}\Lc(\xbf^{(n)}, \etabf^{(n)}), \label{update_a_1}\\
\bar{\etabf}^{(n)} &= \big[\etabf^{(n)}+\alpha\nabla_{\etabf^{(n)}}\Lc(\xbf^{(n)}, \etabf^{(n)})\big]^{+}, \label{update_eta_1}\\
\xbf^{(n+1)}&= \xbf^{(n)}-\alpha\nabla_{\bar{\xbf}^{(n)}}\Lc(\bar{\xbf}^{(n)}, \bar{\etabf}^{(n)}), \label{update_a_2}\\
\etabf^{(n+1)}&= \big[\etabf^{(n)}+\alpha\nabla_{\!\bar{\etabf}^{(n)}}\Lc(\bar{\xbf}^{(n)}, \bar{\etabf}^{(n)})\big]^{+} \label{update_eta_2}
\end{align}
where $\bar{\xbf}^{(n)}$ and $\bar{\etabf}^{(n)}$ are the intermediate updates in the extra update-and-projection step in \eqref{update_a_1}\eqref{update_eta_1},
 $\alpha$ is the step size, and notation $[\zbf]^{+}\triangleq$ $[[z_1]^+, \ldots, [z_n]^+]^T$
with $[z_i]^{+} \triangleq$ $\max\{z_i, 0\}$,
for  $\zbf \in \mathbb{R}^n$.

The gradient $\nabla_{\xbf}\Lc(\xbf,\etabf)$ can be denoted as $\nabla_{\xbf}\Lc(\xbf,\etabf)=[\nabla_{\xbf_1}\Lc(\xbf, \etabf)^T, \cdots, \nabla_{\xbf_G}\Lc(\xbf,\etabf)^T]^T$. From \eqref{Lagrangian},   we obtain
\begin{align}\label{grad_L_a}
\nabla_{\xbf_{i}}\Lc(\xbf, \etabf) & = 2\Abf_{i}^{T}\Abf_{i}\xbf_{i}+2\bigg(\sum_{j\neq i}\sum_{k=1}^{K_j}\gamma_{jk}\eta_{jk}\Fbf_{ijk}\bigg)\xbf_{i}\nonumber\\
&\qquad -2\bigg(\sum_{k=1}^{K_i}\eta_{ik}\Fbf_{iik}\bigg)^{T}\ybf_{i},\,\,\,\, i\in\Gc.
\end{align}
Also from \eqref{Lagrangian}, we obtain the gradient $\nabla_{\etabf}\Lc(\xbf, \etabf)$ as
\begin{align}\label{grad_L_eta}
\nabla_{\etabf}\Lc(\xbf, \etabf) = [\phibf^{T}_{1}(\xbf), \cdots, \phibf^{T}_{G}(\xbf)]^T.
\end{align}
Substituting the expressions in \eqref{grad_L_a} and \eqref{grad_L_eta} into (\ref{update_a_1})--(\ref{update_eta_2}), we obtain the closed-form updates for $\xbf^{(n+1)}$ and $\etabf^{(n+1)}$.

For a monotone variational inequality problem with operator being $L$-Lipschitz continuous,  the extragradient algorithm is guaranteed to converge to the optimal solution, provided that the step size satisfies $\alpha < 1/L$ \cite{Facchinei&Pang:book2003}.
Unfortunately, it is difficult to determine  Lipschitz constant $L$ for  $g(\ubf)$ in our problem.
To overcome this difficulty,
instead of a constant step size $\alpha$, we adopt an adaptive strategy based on the prediction-correction
procedure  \cite{Khobotov:USSR1987,Marcotte:INFORM1991} to adaptively set the step size $\alpha^{(n)}$ for each iteration.

Given a fixed step size $\alpha$, the prediction-correction procedure contains two steps at iteration $n+1$:
\begin{enumerate}
\item Prediction: Obtain $\bar{\ubf}^{(n)}\triangleq(\bar{\xbf}^{(n)}, \bar{\etabf}^{(n)})$ from \eqref{update_a_1} and \eqref{update_eta_1} with fixed $\alpha$. Compute  $d^{(n)}_{\ubf} \triangleq \|\bar{\ubf}^{(n)}\!-\ubf^{(n)}\|$
and $d^{(n)}_{g} \triangleq \|g(\bar{\ubf}^{(n)})\!-g(\ubf^{(n)})\|$.
     Compute step size $\hat{\alpha}^{(n)}$ $=c \frac{d^{(n)}_{\ubf}}{d^{(n)}_{g}}$, where $c \in (0,1)$ is a constant.
\item  Correction: Set  ${\alpha}^{(n)}=\min\{\alpha,\hat{\alpha}^{(n)}\}$  for iteration $n+1$ to update  $\xbf^{(n+1)}$ and $\etabf^{(n+1)}$ in \eqref{update_a_1}--\eqref{update_eta_2}.
\end{enumerate}

We now show that the prediction-correction procedure guarantees
the extragradient algorithm converges to the saddle point
$\ubf^{o}$ of problem \eqref{var_ineq}.
If we replace $\alpha$ by $\alpha^{(n)}$   in  (\ref{update_a_1})--(\ref{update_eta_2}) of the extragradient algorithm, then for any step size sequence $\{\alpha^{(n)}\}$, the following holds \cite{Khobotov:USSR1987,Marcotte:INFORM1991}
 \begin{align}
& \hspace*{-.8em}\|\ubf^{(n+1)}\!-\!\ubf^{o}\|^{2}\!\nn\\[-.5em]
& \hspace*{-.6em} \leq\! \|\ubf^{(n)}\!-\!\ubf^{o}\|^{2} -\|\bar{\ubf}^{(n)}\!-\!\ubf^{(n)}\|^{2}\bigg(\!1\!-\!\bigg(\!\alpha^{(n)}\frac{d^{(n)}_{g}}{d^{(n)}_{\ubf}}\bigg)^{\!\!2}\bigg).
\label{extragradient_converg2}
 \end{align}
If we set  $\alpha^{(n)}$ as in the correction step of the above prediction-correction procedure, then
$\|\ubf^{(n+1)}\!-\!\ubf^{o}\|<\! \|\ubf^{(n)}\!-\!\ubf^{o}\|$ and $\{\ubf^{(n)}\}$ move towards  the saddle point $\ubf^{o}$ of problem \eqref{var_ineq}.  It follows that the algorithm converges to the optimal solution to
$\Pc_{1\text{SCA}}^{\text{r}}(\ybf)$ in each SCA iteration.

Overall, the  ESCA algorithm to solve $\Pc_1$ is summarized in Algorithm~\ref{alg:ESCA}.
The main computational complexity of ESCA lies in computing the gradients  $\nabla_{\xbf}\Lc(\xbf, \etabf)$  using \eqref{grad_L_a} and $\nabla_{\etabf}\Lc(\xbf, \etabf)$ in \eqref{grad_L_eta}.
At each extragradient iteration,
the related matrix-vector computation
for $\nabla_{\xbf}\Lc(\xbf, \etabf)$
requires $\sum_{i=1}^{G}\left(\sum_{j=1}^{G} 4K_jK_i^2 + 32K_i^2\right)$ flops,
and that for $\nabla_{\etabf}\Lc(\xbf, \etabf)$ requires $\sum_{i=1}^{G} \left(\sum_{j=1}^{G}(16K_j^2 + 8K_j)K_i - 8K_i^2\right)$  flops. Note that ESCA consists of two layers of
iterations:  the outer-layer SCA and the inner-layer extragradient-based algorithm to solve each $\Pc_{1\text{SCA}}({\bf{v}})$.
The
number of iterations at the two layers for convergence depend on the system parameters $N$ and $K_\text{tot}$.
 From our experiments, for $N= 100\sim 500$ antennas and $K_\text{tot}= 15\sim 60$ users,
 it typically takes $20\sim 200$ iterations for the inner-layer extragradient algorithm to converge at each SCA iteration,
 and the outer layer typically takes  $2\sim 60$ iterations to converge.

\begin{remark}
As mentioned earlier,
BPA consists of only one update-and-projection step at each iteration (similar to \eqref{update_a_1}\eqref{update_eta_1}),
but requires the operator $g(\ubf)$ to be strongly monotone for convergence.
If  $g(\ubf)$ is only monotone,
the updated point (the same as $\bar{\ubf}^{(n)}$ in \eqref{update_a_1}\eqref{update_eta_1} in iteration $n+1$) may move away from, instead of closer to, the optimal point
$\ubf^{o}$ at each iteration.
 Thus, the updating procedure may not converge over iterations.
In contrast,
the extragradient algorithm
adds an extra update-and-projection step as in \eqref{update_a_2}\eqref{update_eta_2} at each iteration.
This extra step can ensure
convergence
for a monotone operator $g(\ubf)$.
Specifically,
it is shown in \cite{Facchinei&Pang:book2003} that in iteration $n+1$,
after obtaining $\bar{\ubf}^{(n)}$ at the first update-and-projection step in \eqref{update_a_1}\eqref{update_eta_1},
 a hyperplane $\Hc^{(n)} \triangleq \{\ubf \,|\, g(\bar{\ubf}^{(n)})^T(\ubf-\bar{\ubf}^{(n)})= 0\}$
can be constructed at $\bar{\ubf}^{(n)}$ with normal vector $g(\bar{\ubf}^{(n)})$.
For the monotone operator $g(\ubf)$,
it is proven that
this hyperplane $\Hc^{(n)}$ separates the point $\ubf^{(n)}$
 and the optimal point $\ubf^{o}$,
 where $\ubf^{o}$ is in the halfspace  in the direction $-g(\bar{\ubf}^{(n)})$.
 The second update-and-projection step in \eqref{update_a_2}\eqref{update_eta_2} then utilizes
 $-g(\bar{\ubf}^{(n)})$
 to update $\ubf^{(n+1)}$.
 This ensures that
 $\ubf^{(n+1)}$ move towards the optimal point $\ubf^{o}$
 and thus is closer to $\ubf^{o}$
 than $\ubf^{(n)}$ is.
\end{remark}

\begin{algorithm}[t]
\caption{ESCA Algorithm to Solve $\Pc_{1}$}
\label{alg:ESCA}
\begin{algorithmic}[1]
\STATE \textbf{Initialization:} Set  feasible initial point $\ybf^{(0)}$. Set  $\alpha$ and $c$. Set  $l = 0$.
\REPEAT
\STATE // solve $\Pc_{1\text{SCA}}^{\text{r}}(\ybf^{(l)})$
\STATE \textbf{Initialization:} $\xbf^{(0)} = \ybf^{(l)}$, $\etabf^{(0)}= {\bf{0}}$, $n = 0$.
\REPEAT
\STATE Compute $\bar{\xbf}^{(n)}$ and $\bar{\etabf}^{(n)}$ by \eqref{update_a_1}\eqref{update_eta_1} using $\alpha$.
\STATE Compute $g(\bar{\ubf}^{(n)})$ by \eqref{grad_L_a}\eqref{grad_L_eta}.
\STATE Compute $d^{(n)}_{\ubf}$ and $d^{(n)}_{g}$.
\STATE \textit{Prediction:} Compute $\hat{\alpha}^{(n)} = cd^{(n)}_{\ubf}/d^{(n)}_{g}$.
\STATE \textit{Correction:} Update step size ${\alpha}^{(n)}=\min\{\alpha,\hat{\alpha}^{(n)}\}$.
\IF {${\alpha}^{(n)}=\alpha$}
       \STATE   Update $\xbf^{(n+1)}$ and $\etabf^{(n+1)}$ by \eqref{update_a_2}\eqref{update_eta_2}.
   \ELSE
        \STATE Update $\xbf^{(n+1)}$ and $\etabf^{(n+1)}$ by \eqref{update_a_1}--\eqref{update_eta_2} using ${\alpha}^{(n)}$.
   \ENDIF
\STATE $n \leftarrow n+1$.
\UNTIL{convergence}
\STATE Set $\ybf^{(l+1)} = \xbf^{(n)}$. Set $l\leftarrow l+1$.
\UNTIL{convergence}
\end{algorithmic}
\end{algorithm}

\subsection{Initialization for ESCA}
\label{subsec:Ini_ESCA}

A  challenge in using SCA  to solve $\Pc_1$
is that it requires a feasible initial
point satisfying the SINR constraint \eqref{Prb:P_2_constraint}
as $\vbf^{(0)}$ for $\Pc_{1\text{SCA}}({\bf{v}})$
(equivalently $\ybf^{(0)}$ for $\Pc_{1\text{SCA}}^{\text{r}}(\ybf))$.
It is necessary to generate a feasible initial point with a low computational complexity.
Furthermore,
a good initial point closer to the (locally) optimal point of $\Pc_1$
 could accelerate the convergence.
Below, we consider two initialization methods for ESCA.

\emph{1) Extragradient-based initialization method (EIM)}:
Based on the extragradient-based algorithm in Section~\ref{subsec:ESCA_B},
we propose EIM as follows.
EIM
generates a feasible point by solving the following feasibility problem
\begin{align}
\Pc_{1\text{fea}}: & \,\, \text{Find} \,\, \{\xbf\}  \nn\\
&  \text{s.t.} \,\,\frac{\xbf_{i}^{T}\Fbf_{iik}\xbf_{i}}{\sum_{j\neq i}\xbf_{j}^{T}\Fbf_{jik}\xbf_{j}+\sigma^{2}}\geq \gamma_{ik},\, k\in \Kc_{i}, i\in \Gc
\label{3fea}
\end{align}
where
$\xbf_{i}$ and $\Fbf_{jik}$ are defined at the beginning of
Section~\ref{subsec:dual_saddle},
and constraint \eqref{3fea} is an equivalent real representation of constraint \eqref{Prb:P_2_constraint}
in $\Pc_1$ based on $|\abf_{j}^{H}\fbf_{jik}|^{2}=\xbf_{j}^{T}\Fbf_{jik}\xbf_{j}$.

We solve $\Pc_{1\text{fea}}$
by  applying the extragradient method with the adaptive step size
proposed in Section~\ref{subsec:ESCA_B}.
Specifically, the Lagrangian of $\Pc_{1\text{fea}}$ is given by
$\widetilde{\Lc}(\xbf, \widetilde{\bm{\eta}})=\sum_{i=1}^{G}\widetilde{\phibf}_{i}^{T}(\xbf)\widetilde{\bm{\eta}}_{i}$,
where
$\widetilde{\eta}_{ik}$ is the dual variable for constraint \eqref{3fea},
$\widetilde{\bm{\eta}}\triangleq[\widetilde{\bm{\eta}}_{1}^T, \cdots, \widetilde{\bm{\eta}}_{G}^T]^T$
with $\widetilde{\bm{\eta}}_{i}\triangleq[\widetilde{\eta}_{i1}, \cdots, \widetilde{\eta}_{iK_{i}}]^T$,
and
 $\widetilde{\phibf}_{i}(\xbf)\triangleq  [\widetilde{\phi}_{i1}(\xbf),  \ldots, \widetilde{\phi}_{iK_{i}}(\xbf)]^T$
with
$\widetilde{\phi}_{ik}(\xbf)$ $\triangleq $
$\gamma_{ik}\sum_{j\neq i}\xbf_{j}^{T}\Fbf_{jik}\xbf_{j}$ $-\xbf_{i}^{T}\Fbf_{iik}\xbf_{i}  +\gamma_{ik}\sigma^2$,
for $k\in \Kc_{i}, i\in \Gc$.
The gradient  $\nabla_{\xbf_i}\widetilde{\Lc}(\xbf,\widetilde{\etabf})$, for $i\in \Gc$,
is given by
\begin{align}
\nabla_{\xbf_{i}}\widetilde{\Lc}(\xbf, \widetilde{\etabf}) &\! = 2\Big(\!\sum_{j\neq i}\sum_{k=1}^{K_j}\gamma_{jk}\widetilde{\eta}_{jk}\Fbf_{ijk}\!\Big)^{\!\!T}\!\xbf_{i}
 \!-2\Big(\!\sum_{k=1}^{K_i}\widetilde{\eta}_{ik}\Fbf_{iik}\!\Big)^{\!\!T}\!\xbf_{i}. \nn
\end{align}
Similar to \eqref{grad_L_eta},
the gradient $\nabla_{\widetilde{\etabf}}\widetilde{\Lc}(\xbf, \widetilde{\etabf})$
is given by
$\nabla_{\widetilde{\etabf}}\widetilde{\Lc}(\xbf, \widetilde{\etabf}) = [\widetilde{\phibf}^{T}_{1}(\xbf), \cdots, \widetilde{\phibf}^{T}_{G}(\xbf)]^{T}$.
The updating procedure of the extragradient method in \eqref{update_a_1}--\eqref{update_eta_2}
is then applied for solving $\Pc_{1\text{fea}}$,
with $\etabf$, $\nabla_{\xbf}\Lc(\xbf,\etabf)$, and $\nabla_{\etabf}\Lc(\xbf,\etabf)$
being replaced by $\widetilde{\etabf}$, $\nabla_{\xbf}\widetilde{\Lc}(\xbf,\widetilde{\etabf})$, and $\nabla_{\widetilde{\etabf}}\widetilde{\Lc}(\xbf, \widetilde{\etabf})$, respectively.
Furthermore,
the prediction-correction procedure
in Section~\ref{subsec:ESCA_B} is used to set the adaptive step size
at each iteration.

For simplicity, we randomly chose the initial point for EIM. Note that this point may not be feasible for SINR constraint \eqref{Prb:P_2_constraint} in $\Pc_{1}$.
Also,
since
$\Pc_{1\text{fea}}$ is a non-convex optimization problem,
EIM may not be guaranteed to converge
or the terminating point may not be feasible as required for $\vbf^{(0)}$ for $\Pc_{1\text{SCA}}({\bf{v}})$.
Thus, EIM
is served as  a heuristic algorithm.
If the point produced by EIM is infeasible,
we may consider different initial points for EIM until a feasible is obtained.
Our extensive simulation experiments show that EIM converges and provides a feasible initial point with a very high probability.

\emph{2) SDR-based initialization method}:
Similar to \cite{Dong&Wang:TSP2020},
we  can  apply SDR along with Gaussian randomization to   $\Pc_{1}$ to
obtain an approximate solution as
a feasible initial point $\ybf^{(0)}$ to $\Pc_{1\text{SCA}}^{\text{r}}(\ybf)$ for ESCA.
Recall that $\Pc_{1}$ is converted from the original problem $\Pc_{o}$
and is of relatively smaller size ($K_{\text{tot}}$ variables and constraints).
SDR can provide a good initial point
when the problem is of small to moderate size,
leading to fast convergence for ESCA.
However, as $K_{\text{tot}}$ becomes large,
the computational complexity of SDR  will increase significantly,
and the quality of the initial point it computes deteriorates. Thus, SDR for initialization is only suitable for small to moderate problems.

\section{ADMM-Based SCA Algorithm}
\label{sec:ASCA}

In this section, we develop another computationally efficient algorithm based on   ADMM \cite{Boyd&etal:book2011}, which is a different optimization technique from ESCA.  ADMM
has drawn  growing popularity in recent years as a robust and fast
numerical method for solving large-scale problems. We propose an ADMM-based algorithm to solve
each SCA subproblem $\Pc_{1\text{SCA}}(\vbf)$.
Define the auxiliary variables
$d_{jik}$ $\triangleq$ $\abf^{H}_{j}\fbf_{jik}$, for $k\in\Kc_{i},i,j\in\Gc$.
Define
$\dbf$ $\triangleq$ $\left[\dbf^{H}_{11}, \cdots, \dbf^{H}_{GK_G}\right]^{H}$ $\in$ $\mathbb{C}^{GK_{\rm tot}}$,
with $\dbf_{ik}$ $\triangleq$ $\left[d_{1ik}, \cdots, d_{Gik}\right]^{T}$ $\in$ $\mathbb{C}^{G}$.
Then, the problem  $\Pc_{1\text{SCA}}({\bf{v}})$ can be equivalently expressed as
\begin{align}
& \Pc_{1\text{ADMM}}({\bf{v}}):
 \min_{\abf, \dbf}\,\, \sum_{j=1}^{G} \|\Cbf_{j}\abf_{j}\|^2  \nonumber\\
 &\,\,\, \text{s.t.}  \,\, d_{jik} = \abf^{H}_{j}\fbf_{jik},\,\, k\in \mathcal{K}_i, i,j\in \Gc \label{eq_d_a}\\
             &\,\,\, \quad   \gamma_{ik}\!\sum_{j\neq i}\!|d_{jik}|^{2}\!+\!|{\bf{v}}_{i}^{H}\!\fbf_{iik}|^{2}\!+\!\gamma_{ik}\sigma^2 \!-\!2\Re{\{d_{iik}\fbf_{iik}^{H}{\bf{v}}_{i}\}}\! \leq\! 0, \nn\\[-1em]
             & \qquad\qquad\qquad\qquad\qquad\qquad\qquad\qquad   k\in \mathcal{K}_i, i\in \Gc. \label{d_constraints}
\end{align}
Denote the feasible set for the inequality constraint \eqref{d_constraints} by $\Fc$.
Define the indicator function for the set $\Fc$ as
\begin{align}
I_{\mathcal{F}}(\dbf)\triangleq
\begin{cases}
0           & \text{if} \,\,\, \dbf\in \Fc\text{,} \nn\\
\infty           & \text{otherwise}\text{.}\nn
\end{cases}
\end{align}
Then, $\Pc_{1\text{ADMM}}({\bf{v}})$ is  equivalent to the following problem
\begin{align}
\Pc_{\text{1ADMM}}^{\prime}({\bf{v}}): &
 \min_{\abf, \dbf}\,\, \sum_{j=1}^{G} \|\Cbf_{j}\abf_{j}\|^2 +  I_{\mathcal{F}}(\dbf) \nonumber\\
             & \,\, \text{s.t.}  \,\, d_{jik} = \abf^{H}_{j}\fbf_{jik},\,\, k\in \mathcal{K}_i, i,j\in \Gc. \label{a_constraints}
\end{align}
By introducing the auxiliary vector $\dbf$ in constraint \eqref{d_constraints},
we construct the equality-constrained problem $\Pc_{\text{1ADMM}}^{\prime}({\bf{v}})$,
where the objective function contains two separate terms for $\dbf$ and $\abf$ only.
This allows us to apply ADMM to decompose the problem into separate subproblems \cite{Boyd&etal:book2011}.
The augmented Lagrangian of $\Pc_{\text{1ADMM}}^{\prime}({\bf{v}})$ is given by
\begin{align}
\Lc_{\rho}(\dbf, \abf, \qbf)= & \sum_{j=1}^{G} \|\Cbf_{j}\abf_{j}\|^2 +  I_{\mathcal{F}}(\dbf) \nn\\
& +\frac{\rho}{2}\sum_{j=1}^{G}\sum_{i=1}^{G}\sum_{k=1}^{K_i}|d_{jik}-\abf^{H}_{j}\fbf_{jik}+q_{jik}|^{2}
\label{ADMM_Lagrangian}
\end{align}
where  $\rho$ $>$ $0$ is the penalty parameter,
and
$q_{jik}\in\mathbb{C}$ is the dual variable associated with constraint \eqref{a_constraints},
and
$\qbf$ $\triangleq$ $\left[\qbf^{H}_{11}, \cdots, \qbf^{H}_{GK_G}\right]^{H}$,
with $\qbf_{ik}$ $\triangleq$ $\left[q_{1ik}, \cdots, q_{Gik}\right]^{T}$.
We now decompose $\Lc_{\rho}(\dbf, \abf, \qbf)$ into two  subproblems for $\dbf$ and $\abf$ separately,
and update $\{\dbf,\abf,\qbf\}$ alternatively.
Our proposed ADMM-based updating procedure
for solving $\Pc_{\text{1ADMM}}^{\prime}({\bf{v}})$ is given as follows:
\begin{align}
\dbf^{(n+1)}= & \mathop{\arg\min}\limits_{\dbf}\Lc_{\rho}(\dbf, \abf^{(n)}, \qbf^{(n)}), \label{d_update_ADMM}\\
\abf^{(n+1)}= & \mathop{\arg\min}\limits_{\abf}\Lc_{\rho}(\dbf^{(n+1)}, \abf, \qbf^{(n)}), \label{a_update_ADMM}\\
q_{jik}^{(n+1)}= & q_{jik}^{(n)}+\big(d_{jik}^{(n+1)}-\abf^{(n+1)H}_{j}\fbf_{jik}\big) \label{q_update_ADMM}
\end{align}
where $n$ is the iteration index.
Since $\Pc_{\text{1ADMM}}^{\prime}({\bf{v}})$ is convex,
the above ADMM procedure
is guaranteed to converge to the optimal solution of $\Pc_{1\text{ADMM}}({\bf{v}})$
\cite{Boyd&etal:book2011}.

The two main updating steps in \eqref{d_update_ADMM} and \eqref{a_update_ADMM} involve solving two optimization problems.
In the following, we derive the closed-form solutions for the two optimization problems in \eqref{d_update_ADMM} and \eqref{a_update_ADMM},
which leads to fast computation at each iteration.

\subsection{Closed-Form $\dbf$-Update}
\label{sec_ADMM_d_update}
Given $\abf^{(n)}$ and $\qbf^{(n)}$, from \eqref{ADMM_Lagrangian},
the update of $\dbf$ in \eqref{d_update_ADMM} is equivalent to solving the following problem
\begin{align}
\Pc_{\text{d}}({\bf{v}}): &
\min_{\dbf} \sum_{j=1}^{G}\sum_{i=1}^{G}\sum_{k=1}^{K_i}|d_{jik}-\abf^{(n)H}_{j}\fbf_{jik}+q^{(n)}_{jik}|^{2} & {\rm s.t.}\,\, \eqref{d_constraints}.   \nn
\end{align}
Problem $\Pc_{\text{d}}({\bf{v}})$ can be decomposed into $K_{\rm tot}$ convex subproblems,
one for each user $k$ in group $i$ given by
\vspace*{-.5em}
\begin{align}
\Pc_{\text{dsub}}({\bf{v}}):
&\min_{\dbf_{ik}}\,\, \sum_{j=1}^{G}|d_{jik}-e^{(n)}_{1,jik}|^{2}  \nonumber\\
{\rm s.t.}&\,\, e_{2,ik}+\gamma_{ik}\sum_{j\neq i}|d_{jik}|^{2}-2\Re{\{d_{iik}e_{3,ik}\}}\leq 0 \label{d_constraint_2}
\end{align}
where
\begin{align}
e^{(n)}_{1,jik}\!\! \triangleq\! \abf^{(n)H}_{j}\!\fbf_{jik}\!\!- \!q^{(n)}_{jik}, \,
e_{2,ik} \!\!\triangleq \!|\vbf_{i}^{H}\!\fbf_{iik}|^{2}\!\!+\!\gamma_{ik}\!\sigma^2, \,
e_{3,ik} \!\!\triangleq \!\fbf^{H}_{iik}\!\vbf_{i}.\label{eq_e1}
\end{align}
Problem $\Pc_{\text{dsub}}({\bf{v}})$ is a convex QCQP-1 problem,
whose solution may be derived in closed-form.
In particular,
a problem of similar structure has been considered in \cite{Chen&Tao:ITC2017},
where the closed-form solution is derived in \cite[Appendix A]{Chen&Tao:ITC2017}.
We directly use this result and state the closed-form solution below.
The optimal solution $\dbf^{o}_{ik}$ for
 $\Pc_{\text{dsub}}({\bf{v}})$ is given by
\begin{align}
d^{o}_{jik} =
\begin{cases}
e^{(n)}_{1,iik}+\nu^{o}_{ik}e^{*}_{3,ik}           & \text{if} \,\,\, j = i, \\
\frac{e^{(n)}_{1,jik}}{1+\nu^{o}_{ik}\gamma_{ik}}           & \text{otherwise} \label{eq_d_optimal}
\end{cases}
\end{align}
where $\nu^{o}_{ik}$ $=$ $0$ if $e_{2,ik}$ $+\gamma_{ik}\sum_{j\neq i}|e^{(n)}_{1,jik}|^{2}$ $-2\Re{\{e_{3,ik}e^{(n)}_{1,iik}\}}$ $ \leq 0$;
otherwise, $\nu^{o}_{ik}$ is the unique real positive root of the following  cubic equation
of $\nu_{ik}$:
\begin{align}
e_{2,ik}\!+\!\gamma_{ik}\frac{\sum_{j\neq i}|e^{(n)}_{1,jik}|^{2}}{(1+\nu_{ik}\gamma_{ik})^{2}}\!-\!2\Re{\{e_{3,ik}e^{(n)}_{1,iik}\}}\!-\!2\nu_{ik}|e_{3,ik}|^{2}\! =\! 0.
\label{eq_cubic}
\end{align}
Since the roots of \eqref{eq_cubic} are given by the cubic formula,
 $\nu^{o}_{ik}$ is obtained in closed-form.
For the sake of completeness, the key steps leading to the above solution is provided in Appendix~\ref{append:deriv_closed_d}.

\subsection{Closed-Form $\abf$-Update}
Given $\dbf^{(n+1)}$ and $\qbf^{(n)}$,
the update of $\abf$ in \eqref{a_update_ADMM} is equivalent to solving the following problem
\begin{align}
& \Pc_{\text{a}}({\bf{v}})\!:  \min_{\abf}\! \sum_{j=1}^{G}\! \left(\!\|\Cbf_{j}\abf_{j}\|^2 \!+\!\frac{\rho}{2} \!\sum_{i=1}^{G}\sum_{k=1}^{K_i}|d^{(n+1)}_{jik}\!-\!\abf^{H}_{j}\fbf_{jik}\!+\!q^{(n)}_{jik}|^{2}\!\right)\!. \nn
\end{align}
Problem $\Pc_{\text{a}}({\bf{v}})$
can be decomposed into $G$ subproblems,
one for each group $j$ expressed as
\begin{align}
\Pc_{\text{asub}}({\bf{v}})\!:\! \min_{\abf_{j}} \! \|\Cbf_{j}\abf_{j}\|^2\!+\!\frac{\rho}{2}\!\sum_{i=1}^{G}\sum_{k=1}^{K_i}\!|d^{(n+1)}_{jik}\!-\!\abf^{H}_{j}\fbf_{jik}\!+\!q^{(n)}_{jik}|^{2}.\nn
\end{align}
Problem $\Pc_{\text{asub}}({\bf{v}})$ is an unconstrained convex optimization problem.
Using the first-order optimality condition \cite{Boyd&Vandenberghe:book2004},
we obtain
the closed-form solution to $\Pc_{\text{asub}}({\bf{v}})$  as
\begin{align}
\abf^{(n+1)}_{j}= & \frac{\rho}{2}\left(\Cbf^{H}_{j}\Cbf_{j}+\frac{\rho}{2}\sum_{i=1}^{G}\sum_{k=1}^{K_i}\fbf_{jik}\fbf^{H}_{jik}\right)^{-1} \nn\\
& \cdot\sum_{i=1}^{G}\sum^{K_i}_{k=1}\left(d^{(n+1)}_{jik}+q^{(n)}_{jik}\right)^{*}\fbf_{jik}.
\label{eq_a_cf}
\end{align}

We summarize the ASCA algorithm in Algorithm~\ref{alg:ASCA}.
The main computational complexity of ASCA lies in
computing
$\{e^{(n)}_{1,jik}\}$  in \eqref{eq_e1} for $k\in\Kc_{i}, i,j\in\Gc$
and $\{\abf^{(n+1)}_{j}\}$ in \eqref{eq_a_cf} for $j\in\Gc$ at each ADMM iteration.
Note that computing $\abf^{(n+1)}_{j}$  involves a
matrix inversion  with a complexity of $O(K^{3}_{j})$.
However,
the matrix only depends on the channel vectors, and thus the matrix inversion
only needs to be computed once
at the beginning of ASCA.
Thus for each ADMM iteration, only matrix-vector multiplications are involved.
At each ADMM iteration,
the related matrix-vector computation
of $\{e^{(n)}_{1,jik}\}$
requires $12\left( \sum_{i=1}^{G} K_i\right)^2 + 2G\sum_{i=1}^{G}K_i$ flops,
and that of $\{\abf^{(n+1)}_{j}\}$ requires $6\left( \sum_{i=1}^{G} K_i\right)^2 + 2\sum_{i=1}^{G}\left(6 K_i^2 + G K_i\right)$  flops. There are two layers of iterations in ASCA:  the outer-layer SCA and the inner-layer ADMM to solve each $\Pc_{1\text{SCA}}({\bf{v}})$.
The convergence speed depends on the system parameters $N$ and $K_\text{tot}$.
 From our experiments, for $N= 100\sim 500$ antennas and $K_\text{tot}= 15\sim 60$ users,
 it typically takes $50\sim 150$ iterations for the inner-layer ADMM-based algorithm to converge at each SCA iteration,
 and the outer layer typically takes  $2\sim 90$ iterations  to converge.

 In Section~\ref{sec:simulations}, we will provide the simulation study to compare the convergence, performance, and the computational time of ESCA and ASCA.

\begin{algorithm}[t]
\caption{ASCA Algorithm to Solve $\Pc_{1}$}
\label{alg:ASCA}
\begin{algorithmic}[1]
\STATE \textbf{Initialization:} Set feasible initial point $\vbf^{(0)}$. Set $\rho$; $l=0$.
\REPEAT
\STATE // solve $\Pc_{1\text{ADMM}}(\vbf^{(l)})$
\STATE \textbf{Initialization:} $\abf^{(0)} = \vbf^{(l)}$, $\dbf^{(0)} = {\bf{0}}$, $\qbf^{(0)} = {\bf{0}}$, $n = 0$.
\REPEAT
\STATE Compute $\dbf^{(n+1)}$ by \eqref{d_update_ADMM} with $\abf^{(n)}$ and $\qbf^{(n)}$.
\STATE Compute $\abf^{(n+1)}$ by \eqref{a_update_ADMM} with $\dbf^{(n+1)}$ and $\qbf^{(n)}$.
\STATE Compute $\qbf^{(n+1)}$ by \eqref{q_update_ADMM} with $\dbf^{(n+1)}$ and $\abf^{(n+1)}$.
\STATE $n \leftarrow n+1$.
\UNTIL{convergence}
\STATE Set $\vbf^{(l+1)} = \abf^{(n)}$. Set $l\leftarrow l+1$.
\UNTIL{convergence}
\end{algorithmic}
\end{algorithm}

\subsection{Initialization for ASCA}
\label{subsec:Ini_ASCA}

As discussed earlier in Section~\ref{subsec:Ini_ESCA},
a feasible initial point is required by SCA to solve $\Pc_{1}$.
Below we propose an ADMM-based initialization method (AIM).

Using the ADMM-based algorithm above,
we propose AIM  for  ASCA.
The feasible point  is  computed by solving the following feasibility problem
\begin{align}
\Pc_{1\text{fea}}^{\prime}: \,\,\text{Find} \,\, \{\abf\} \quad \text{s.t.} \, \, \eqref{Prb:P_2_constraint}. \nn
\end{align}
Using the auxiliary variables
$d_{jik}$ $=$ $\abf^{H}_{j}\fbf_{jik}$, for $k\in\Kc_{i},i,j\in\Gc$, defined at the beginning of
Section~\ref{sec:ASCA},
$\Pc_{1\text{fea}}^{\prime}$ is equivalently expressed as
\begin{align}
&\Pc_{1\text{feaADMM}}^{\prime}: \,\,\text{Find} \,\, \{\abf,\dbf\} \nn\\
& \qquad\quad\quad\,\, \text{s.t.} \, \, d_{jik} = \abf^{H}_{j}\fbf_{jik},\,\, k\in \mathcal{K}_i, i,j\in \Gc \nn\\
& \qquad\quad\quad\quad\,\,\,\, \frac{|d_{iik}|^{2}}{\sum_{j\neq i}|d_{jik}|^{2}+\sigma^2}\geq \gamma_{ik}, \, k\in \mathcal{K}_i, i\in \Gc.
\label{Prb:P_1fea_constraint}
\end{align}
Denote the feasible set for constraint \eqref{Prb:P_1fea_constraint}
by $\widetilde{\mathcal{F}}$.
The augmented Lagrangian of $\Pc_{1\text{feaADMM}}^{\prime}$ is given by
\begin{align}
\!\!\!\widetilde{\Lc}_{\tilde{\rho}}(\dbf, \abf, \tqbf)\!= \! I_{\widetilde{\mathcal{F}}}(\dbf)\!+\!\frac{\tilde{\rho}}{2}\!\sum_{j=1}^{G}\!\sum_{i=1}^{G}\!\sum_{k=1}^{K_i}\!|d_{jik}\!-\!\abf^{H}_{j}\fbf_{jik}\!+\!\tq_{jik}|^{2} \label{ADMM_fea_lagrangian}
\end{align}
where  indicator  function $I$, penalty parameter $\tilde{\rho}$, dual variable $\tq_{jik}$
are defined similarly as those in \eqref{ADMM_Lagrangian}.
Similar to the ADMM updating procedures in \eqref{d_update_ADMM}--\eqref{q_update_ADMM},
based on \eqref{ADMM_fea_lagrangian},
the AIM updating procedure for solving  $\Pc_{1\text{feaADMM}}^{\prime}$ is
given as follows:  At iteration $n+1$,
\begin{align}
\dbf^{(n+1)}= & \mathop{\arg\min}\limits_{\dbf\in\widetilde{\mathcal{F}}}\!
\sum_{j=1}^{G}\!\sum_{i=1}^{G}\!\sum_{k=1}^{K_i}\!|d_{jik}\!-\!\abf^{(n)H}_{j}\fbf_{jik}\! + \!\tq^{(n)}_{jik}|^{2}, \label{d_update_AIM}\\
\abf^{(n+1)}= & \mathop{\arg\min}\limits_{\abf}\!\sum_{j=1}^{G}\!\sum_{i=1}^{G}\!\sum_{k=1}^{K_i}\!|d^{(n+1)}_{jik}\!-\!\abf^{H}_{j}\fbf_{jik}\!+\!\tq^{(n)}_{jik}|^{2}, \label{a_update_AIM}\\
\tq_{jik}^{(n+1)}= & \tq_{jik}^{(n)}+\big(d_{jik}^{(n+1)}-\abf^{(n+1)H}_{j}\fbf_{jik}\big). \label{q_update_AIM}
\end{align}
The updating steps in \eqref{d_update_AIM} and \eqref{a_update_AIM} can be derived in closed-form,
which are provided in Appendix~\ref{append:derivation_AIM}.

The initial point for AIM is randomly chosen,
which may not be feasible for SINR constraint \eqref{Prb:P_1fea_constraint}
in $\Pc_{1\text{feaADMM}}^{\prime}$.
However, if AIM converges, the terminating point will satisfy SINR constraint \eqref{Prb:P_1fea_constraint},
and the produced point $\abf$ for $\vbf^{(0)}$ is feasible to $\Pc_{1\text{SCA}}({\bf{v}})$.
Note that since
$\Pc_{1\text{fea}}^{\prime}$ is a non-convex optimization problem,
the ADMM procedure for AIM described above may not be guaranteed to converge.
Thus, AIM
is served as  a heuristic algorithm.
Our extensive simulation studies show that AIM converges to a feasible point with a very high probability.

Besides AIM,
We can also use the SDR-based initialization method discussed in Section~\ref{subsec:Ini_ESCA}
 for initialization  of ASCA,
where
a feasible initial point $\vbf^{(0)}$ for ASCA
is obtained by applying
SDR along with Gaussian randomization to solve
$\Pc_{1}$.

\section{Scaling Scheme for the MMF Problem}
\label{sec:duality}
In this section, with the proposed ESCA and ASCA for the QoS problem,
we propose an efficient scheme to obtain a solution to the MMF problem $\Sc_{o}$.

We transform the original MMF problem $\Sc_{o}$ into the following equivalent problem
\begin{align}
\Sc^{\prime}_{o}(\gammabf, P):  \max_{{\bf{w}},t} & \ \  t \nonumber \\
\text{s.t.}& \ \   \text{SINR}_{ik}\geq t\gamma_{ik},\,\, k\in \Kc_i, i\in \Gc \nn\\
&\ \  \sum_{i=1}^{G} \|\wbf_i\|^2\leq P \nn
\end{align}
where
vector $\gammabf$ contains all
SINR targets $\{\gamma_{ik}\}$ of all users in all groups.
Furthermore, we parameterize the QoS problem $\Pc_{o}$  as $\Pc_{o}(\gammabf)$.
It has been shown that $\Sc^{\prime}_{o}(\gammabf, P)$ and
$\Pc_{o}(t\gammabf)$ are the inverse problems to each other \cite{Karipidis&etal:TSP2008}.
Specifically,
denote the maximum objective value of $\Sc^{\prime}_{o}(\gammabf, P)$
by $t^{o}=\Sc^{\prime}_{o}(\gammabf, P)$.
Let the minimum power objective value of $\Pc_{o}(\gammabf^{\prime})$ be $P = \Pc_{o}(\gammabf^{\prime})$
for some $\gammabf^{\prime}$.
Then, we have the following inverse relation:
\begin{align}
t^{o} =  \Sc^{\prime}_{o}(\gammabf, \Pc_{o}(t^{o}\gammabf)),  \quad\,\,  P = \Pc_{o}(\Sc^{\prime}_{o}(\gammabf, P)\gammabf).
\label{duality_iterative}
\end{align}
Based on this relation, in the literature, the MMF problem $\Sc^{\prime}_{o}(\gammabf, P)$
is typically solved via iteratively solving the QoS problem $\Pc_{o}(t\gammabf)$
along with a bi-section search over $t$ until the transmit power objective in $\Pc_{o}(t\gammabf)$
is equal to
$P$ \cite{Karipidis&etal:TSP2008, Ottersten&etal:TSP14, Christopoulos&etal:SPAWC15, Chen&Tao:ITC2017, Dong&Wang:TSP2020}.
However, this approach is computationally expensive.
As mentioned at the end of Section~\ref{subsec:SCA_prelimi},
existing works use either SDR  or SCA to compute a solution to $\Pc_{o}(t\gammabf)$,
where the
second-order IPM is used to solve either relaxed or convexified approximate problem.
As a result,
iteratively solving $\Pc_{o}(t\gammabf)$
incurs high computational complexity not suitable for large-scale problems.
Using the optimal beamforming structure $\wbf^{o}_{i}$ in (\ref{optimal_QoS_structure})
can
 significantly reduce the required computation in the above approach, where $\Pc_{o}(t\gammabf)$ can be converted to a much smaller weight optimization problem as in $\Pc_1$. Nonetheless, resorting to the additional layer of  iterations to solve the QoS problem  is still  computationally costly for the MMF problem as compared with the QoS problem itself, especially for large-scale problems \cite{Dong&Wang:TSP2020}.

\begin{algorithm}[t]
\caption{The Closed-Form Scaling Scheme for $\Sc_o(\gammabf, P)$}
\label{alg:MMF}
\begin{algorithmic}[1]
\STATE Solve $\Pc_{o}(\gammabf)$ and attain solution $\wbf^{\sssQ}(\gammabf)$.
\STATE Compute $P^{\sssQ}(\gammabf)=\sum_{i=1}^{G} \|\wbf^{\sssQ}_i(\gammabf)\|^2$.
\STATE Compute $\wbf^{\text{s}}(\gammabf, P)=$ $\sqrt{\frac{P}{P^{\sssQ}(\gammabf)}}\wbf^{\sssQ}(\gammabf)$ as the solution to $\Sc_o(\gammabf, P)$.
\end{algorithmic}
\end{algorithm}

To avoid the additional iterative procedure,
we develop a closed-form scaling scheme for
finding a solution to $\Sc_{o}$ directly.
Specifically,
we
first obtain the solution to $\Pc_{o}(\gammabf)$ by solving the smaller weight optimization problem $\Pc_{1}$
using either ESCA or ASCA proposed in Sections~\ref{sec:ESCA} and \ref{sec:ASCA}.
Then, we scale this solution to $\Pc_{o}(\gammabf)$ to obtain
a solution to $\Sc^{\prime}_{o}(\gammabf, P)$,
 such that the transmit power budget
 $P$ is met.
Parameterize $\Sc_o$ as $\Sc_o(\gammabf, P)$.
This proposed scaling scheme is summarized in  Algorithm~\ref{alg:MMF}. We
show below that this scaling scheme
provides a feasible solution to $\Sc_{o}(\gammabf, P)$, and we also bound its performance.

\begin{proposition}
\label{thm:duality}
Let $\wbf^{\sss \text{Q}}(\gammabf)$
be a feasible beamforming solution to $\Pc_{o}(\gammabf)$ produced by a given algorithm,
with the achieved objective value denoted by $P^{\sss \text{Q}}(\gammabf)$.
Define
$I^{\sssQ}_{ik}(\gammabf) \triangleq \sum_{j\neq i}|\hbf_{ik}^{H}\wbf_j^{\sssQ}(\gammabf)|^{2}$.
Then, the scaled beamforming vector $\wbf^{\text{s}}(\gammabf, P)\triangleq$
$\sqrt{\frac{P}{P^{\sssQ}(\gammabf)}}\wbf^{\sssQ}(\gammabf)$ is a feasible solution
to $\Sc_o(\gammabf, P)$;
the corresponding achieved objective value, denoted by
$t^{\text{s}}(\gammabf, P)$, satisfies
\begin{align}
& \frac{P}{P^{\sssQ}(\gammabf)}\min_{i,k}    \frac{{I^{\sssQ}_{ik}(\gammabf)+\sigma^2}}{{\frac{P}{P^{\sssQ}(\gammabf)}I^{\sssQ}_{ik}(\gammabf)+\sigma^2}}
\leq
t^{\text{s}}(\gammabf, P) \nn\\
& \qquad\qquad\quad
\leq
\frac{P}{P^{o}(\gammabf)} \max_{i,k} \frac{I^{\sssQ}_{ik}(\gammabf)+\sigma^{2}}{\frac{P}{P^{\sssQ}(\gammabf)}I^{\sssQ}_{ik}(\gammabf)+\sigma^{2}}
\label{scaling_bound}
\end{align}
where  $P^{o}(\gammabf)$  denotes  the optimal objective value of $\Pc_o(\gammabf)$.
\end{proposition}

\begin{IEEEproof}
See Appendix~\ref{append:proof_thm_1}.
\end{IEEEproof}

Note that the tightness
of the lower and upper bounds for $t^{\text{s}}(\gammabf, P)$ in \eqref{scaling_bound}
depends on transmit power $P^{\sssQ}(\gammabf)$,
which is
 obtained by a given algorithm for $\Pc_{o}(\gammabf)$ with solution $\wbf^{\sssQ}(\gammabf)$.
If
the power budget $P$ for the MMF problem $\Sc_o(\gammabf, P)$ is more than
the optimal power value for $\Pc_{o}(\gammabf)$,
\ie
$P\geq P^{o}(\gammabf)$,
and
the algorithm provides $P^{\sssQ}(\gammabf) = P$,
then $\wbf^{\text{s}}(\gammabf, P)=\wbf^{\sssQ}(\gammabf)$,
and
the bounds in \eqref{scaling_bound} in this case
are simplified
to $1\leq t^{\text{s}}(\gammabf, P) \leq \frac{P}{P^{o}(\gammabf)}$.
Note that, often for the given algorithm, the solution $\wbf^{\sssQ}(\gammabf)$
 results in at least one SINR constraint being attained with equality,
 \ie
 $\text{SINR}_{ik} = \gamma_{ik}$,
  for some $i, k$,
  then we have $t^{\text{s}}(\gammabf, P)=1$.
In a special case when $P^{\sssQ}(\gammabf) = P = P^{o}(\gammabf)$,
we have $t^{\text{s}}(\gammabf, P)=1$,
and also both
the upper and lower bounds become $1$.
In this case,
since $P = P^{o}(\gammabf)$,
from the inverse relation in \eqref{duality_iterative},
we have $t^{o} = 1$.
Thus, $t^{\text{s}}(\gammabf, P) = t^{o} = 1$,
and the bounds in \eqref{scaling_bound} are tight.

Comparing    Algorithm~\ref{alg:MMF} with the iterative bi-section search method
 using \eqref{duality_iterative},
we see that
our proposed closed-form scaling scheme avoids iteratively solving $\Pc_{2}(t\gammabf)$
along with a bi-section search over $t$,
and thereby,
enjoys a significant reduction of computational complexity.
Moreover,
we can directly apply the fast algorithm
ESCA proposed in Section~\ref{sec:ESCA} or ASCA proposed in Section~\ref{sec:ASCA}
along with the optimal structure in \eqref{optimal_QoS_structure}
to provide a solution to $\Pc_{o}(\gammabf)$ with this scaling scheme.
This approach leads to two fast first-order algorithms for solving  $\Sc_{o}(\gammabf, P)$
in large-scale systems.

\begin{remark}
We point out that
our scaling scheme for the multi-group multicast beamforming MMF problem is
different from a similar scaling scheme
proposed for the  MMF problem in a single-group scenario
in \cite{Sadeghi&etal:TWC17}.
Specifically,
the scheme  in \cite{Sadeghi&etal:TWC17}
first applies ZF beamforming for each group to eliminate inter-group interference.
Once the interference is removed, for the equivalent single-group multicast beamforming problem,
\cite{Sadeghi&etal:TWC17}
scales the beamforming solution of the
single-group QoS problem to obtain a feasible solution to the MMF problem.
In our scheme, we directly handle the original multi-group MMF problem containing inter-group interference.
Our scheme scales the solution of the multi-group QoS problem $\Pc_{o}(\gammabf)$
to solve the original MMF problem $\Sc_{o}(\gammabf, P)$.
Note that
the scheme in \cite{Sadeghi&etal:TWC17} has
the lower and upper bounds for the objective value $t$
(similar to $t$ in $\Sc^{\prime}_{o}(\gammabf, P)$)
as $\big[\frac{P}{P^{\sssQ}(\gammabf)}, \frac{P}{P^{o}(\gammabf)}\big]$
with no interference present.
In contrast, for our scheme,
the lower and upper bounds in \eqref{scaling_bound} contain
additional terms w.r.t.\ $I^{\sssQ}_{ik}(\gammabf)$.
Note that
$I^{\sssQ}_{ik}(\gammabf)$ represents the inter-group interference
to user $k$ in group $i$ by solution $\wbf^{\sssQ}(\gammabf)$.
Following this,
 both terms $I^{\sssQ}_{ik}(\gammabf)+\sigma^2$
and $\frac{P}{P^{\sssQ}(\gammabf)}I^{\sssQ}_{ik}(\gammabf)+\sigma^2$ in
\eqref{scaling_bound} are
the interference plus noise term for user $k$ in group $i$,
where the latter is
based on the scaled beamforming solution $\wbf^{\text{s}}(\gammabf, P)$.
These additional terms in the lower and upper bounds in \eqref{scaling_bound}
 represent the minimum and maximum inter-group interference ratios, respectively.
In the special case of single group $G=1$,
the bounds in \eqref{scaling_bound} reduces to $\big[\frac{P}{P^{\sssQ}(\gammabf)}, \frac{P}{P^{o}(\gammabf)}\big]$
in \cite{Sadeghi&etal:TWC17}.
Thus, the bounds in \eqref{scaling_bound} can be viewed as a generalization of
 the bounds in \cite{Sadeghi&etal:TWC17}
from single-group to multi-group settings with inter-group interference.
\end{remark}

\section{Simulation Results}
\label{sec:simulations}
We consider a default setup for downlink multi-group multicast beamforming,
where $G=3$ groups,  $K_{i} = K$ users/group, $\forall~ i\in\Gc$,
and the same SINR target for all users as  $\gamma_{ik}=\gamma=10~\text{dB}$, $\forall k, i$.
The user channels are generated  independently with an identical distribution\ as $\hbf_{ik}\sim\mathcal{CN}({\bf{0}},{\bf{I}})$.
The performance plots are obtained by averaging the results over 100 channel realizations per user.

\begin{figure}[t]
\centering
\hspace*{-.4cm}\includegraphics[width=0.55\textwidth,height=3in]{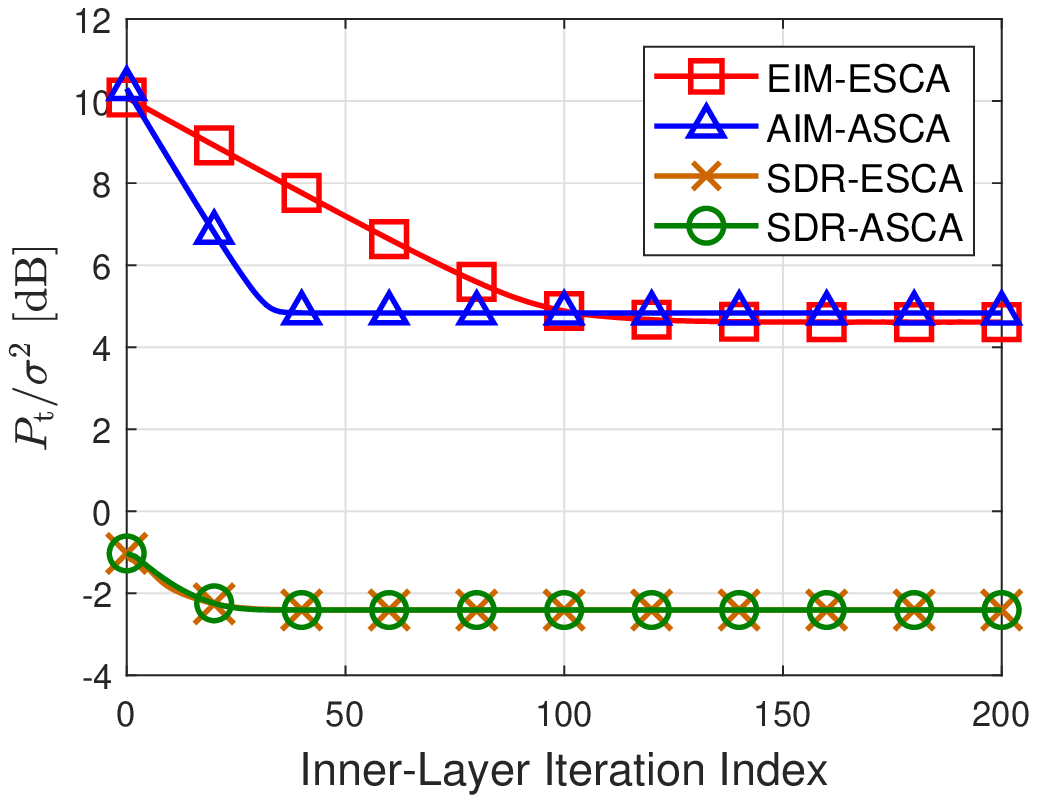}\vspace{-5em}
\caption{Convergence behavior for $\Pc_{o}$: Normalized power objective $P_{\rm t}/\sigma^2$ over the inner-layer iterations at the first outer-layer SCA iteration ($N = 500$, $K = 10$).}\vspace*{.5em}
\label{Fig1:QoS_Convergence_Inner_Power}
\centering
\hspace*{-.4cm}\includegraphics[width=0.55\textwidth,height=3in]{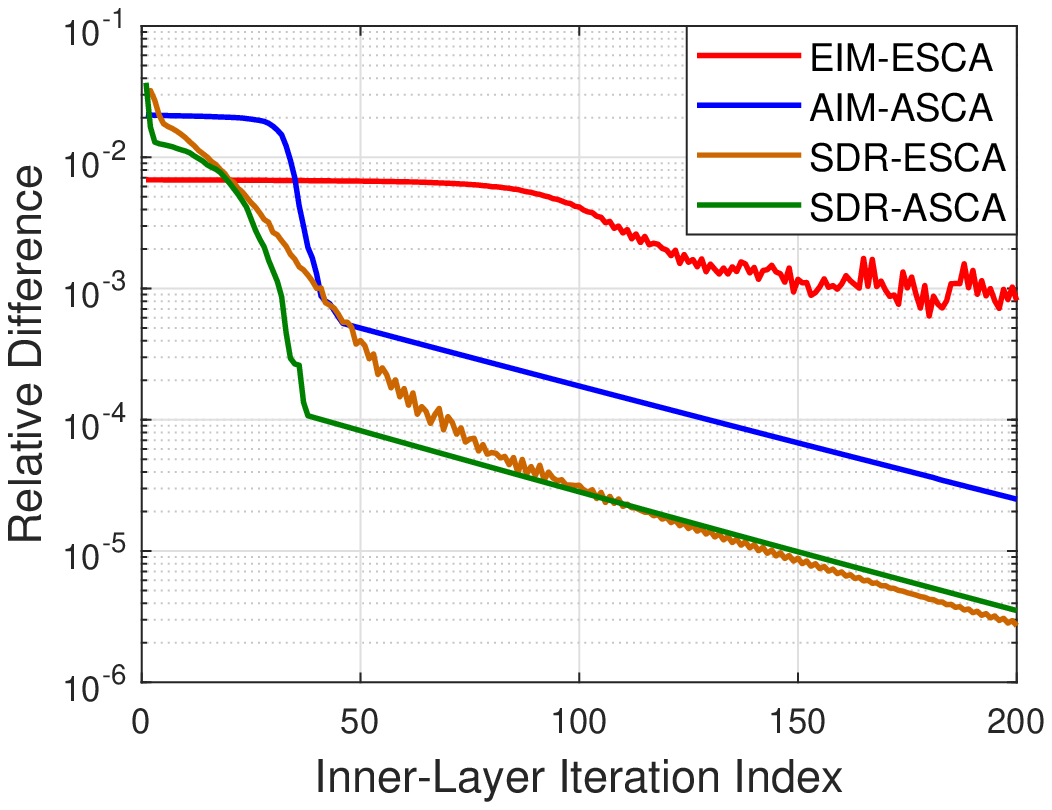}\vspace{-5em}
\caption{Convergence behavior  for $\Pc_{o}$: Relative difference over the inner-layer iterations at the first outer-layer SCA iteration ($N = 500$, $K = 10$).}\vspace*{.5em}
\label{Fig1_2:QoS_Convergence_Inner_RD}
\end{figure}

For QoS problem $\Pc_{o}$,
we consider the proposed two fast algorithms:
ESCA in Algorithm~\ref{alg:ESCA}
and ASCA in Algorithm~\ref{alg:ASCA}.
For ESCA,
we set the step size $\alpha = 0.1$ and the constant $c = 0.8$.
For ASCA,
we set the penalty parameter $\rho = 0.2$.\footnote{We have studied different values of $\rho$ and found $\rho = 0.2$ generally provides overall good  performance and convergence speed for ASCA.}
We consider different initialization methods discussed in Sections~\ref{subsec:Ini_ESCA} and \ref{subsec:Ini_ASCA}
for ESCA and ASCA, respectively.
Therefore, we refer to our algorithms as follows: 1) EIM-ESCA: ESCA with EIM initialization;
2) SDR-ECSA: ESCA with SDR initialization;
3) AIM-ASCA: ASCA with AIM initialization;
4) SDR-ASCA: ASCA with SDR initialization.
Note that each algorithm solves the weight optimization problem $\Pc_{1}$
using the optimal beamforming structure in \eqref{optimal_QoS_structure}.\footnote{Note that in \eqref{optimal_QoS_structure},  the exact expression of ${\bf{R}}(\bm{\lambda}^{o})$ is used and its computation is discussed below \eqref{optimal_QoS_structure}.
One may utilize the asymptotic expression of ${\bf{R}}(\bm{\lambda}^{o})$ obtained in \cite{Dong&Wang:TSP2020}
to simplify the computation of ${\bf{R}}(\bm{\lambda}^{o})$.
Since the fixed-point iterative method \cite{Dong&Wang:TSP2020} for computing  ${\bf{R}}(\bm{\lambda}^{o})$
is computationally inexpensive,
using the asymptotic expression of ${\bf{R}}(\bm{\lambda}^{o})$  only brings a minor reduction of computation
cost and thus is not considered in the simulation.}
Besides the above four algorithms for solving  $\Pc_{1}$,
we also consider the following methods for comparison:
\begin{itemize}
\item Lower Bound for $\Pc_{o}$: Obtained by solving  the relaxed version of $\Pc_{o}$ via SDR.
\item SDR-CSCA~\cite{Dong&Wang:TSP2020}\footnote{Note that we only consider SDR-CSCA as the benchmark for  comparison against our proposed algorithms. This is because  SDR-CSCA is the state-of-the-art method
 with a near-optimal performance and  substantially lower computational complexity
than other existing algorithms in the literature. The comparison of SDR-CSCA and other existing algorithms in both performance and computational complexity  has already been provided in \cite{Dong&Wang:TSP2020}, and adding other algorithms here will not provide additional insight or observation than what has been shown in \cite{Dong&Wang:TSP2020}.}: Solve  $\Pc_{1}$ via SCA using $\Pc_{1\text{SCA}}({\bf{v}})$,
 where each $\Pc_{1\text{SCA}}({\bf{v}})$ is solved by the standard convex solver CVX, which uses IPM. The SDR method is used to generate an initial point.
\end{itemize}

For MMF problem $\Sc_o$,
we consider ESCA and ASCA with the proposed scaling scheme in Algorithm~\ref{alg:MMF}.
They
are denoted as ESCA-Scaling and ASCA-Scaling, respectively.
For comparison,
we also consider the following methods:
\begin{itemize}
\item Upper Bound for $\Sc_{o}$: Obtained by solving the relaxed version of $\Pc_o$ using SDR along with the bi-section search over $t$.
\item ESCA-Bisection: Solve $\Sc_{o}$ via iteratively solving $\Pc_{o}$ along with bi-section search over $t$.  For solving $\Pc_{o}$, SDR-ESCA is used, where the optimal beamforming structure
in \eqref{optimal_QoS_structure} with the asymptotic expression of  ${\bf{R}}(\bm{\lambda}^{o})$ obtained in \cite{Dong&Wang:TSP2020} is applied.
\item ASCA-Bisection: Similar to ESCA-Bisection, except that SDR-ASCA is used to solve $\Pc_{o}$.
\item CSCA-Bisection \cite{Dong&Wang:TSP2020}: Similar to ESCA-Bisection, except that SDR-CSCA is used to solve $\Pc_{o}$.
\item SDR-Bisection \cite{Dong&Wang:TSP2020}: Solve $\Sc_{o}$ via iteratively solving $\Pc_{o}$ along with bi-section search over $t$.  For solving $\Pc_{o}$, SDR along with the Gaussian randomization method is used, where the optimal beamforming structure
in \eqref{optimal_QoS_structure} with the asymptotic expression of  ${\bf{R}}(\bm{\lambda}^{o})$ is applied.
\end{itemize}

\subsection{Convergence Analysis}

In this subsection, we study the convergence behavior of
the two proposed fast algorithms (ESCA and ASCA) for the QoS problem $\Pc_o$.
Note that each algorithm consists of two layers of iterations:
Outer-layer SCA and the inner-layer iterative algorithm to solve each $\Pc_{1\text{SCA}}({\bf{v}})$.
Let $P_{\rm t}$ $=$ $\sum_{i=1}^{G} \|\wbf_i\|^2$ denote the total transmit power.
Fig.~\ref{Fig1:QoS_Convergence_Inner_Power} shows the trajectory of
normalized transmit power $P_{\rm t}/\sigma^2$
over the inner-layer iterations at the first outer-layer SCA iteration.
We set $N = 500$ and $K = 10$.
We observe that SDR-ESCA and SDR-ASCA converge faster and result in lower  $P_{\rm t}/\sigma^2$
 than EIM-ESCA and AIM-ASCA. This shows that the SDR initialization method provides
 a better initial point than the other initialization methods.
 Between SDR-ESCA and SDR-ASCA, we observe a similar convergence rate.
Comparing EIM-ESCA and AIM-ASCA,
we see that
AIM-ASCA converges faster than EIM-ESCA.
Next, we consider the convergence behavior of the relative difference of the optimization variable for each algorithm. Specifically, we consider
the normalized difference
 $\frac{\|\xbf^{(n+1)}-\xbf^{(n)}\|}{\|\xbf^{(n+1)}\|}$
in two consecutive inner-layer iterations of ESCA in Algorithm~\ref{alg:ESCA}, and similarly,  $\frac{\|\abf^{(n+1)}-\abf^{(n)}\|}{\|\abf^{(n+1)}\|}$ for ASCA in Algorithm~\ref{alg:ASCA}.
Fig.~\ref{Fig1_2:QoS_Convergence_Inner_RD}
shows these relative differences for different algorithms
 at the first outer-layer SCA iteration
for $N = 500$ and $K = 10$.
Again, we observe that
the SDR initialization method results in a faster convergence than the other initialization methods.
Both SDR-ESCA and SDR-ASCA reach a relative difference of $10^{-3}$ within 50 iterations,
with SDR-ASCA converging slightly faster than SDR-ESCA.
Comparing EIM-ESCA and AIM-ASCA,
we see that
AIM-ASCA provides a faster convergence than EIM-ESCA,
reaching a relative difference of $10^{-3}$ within 50 iterations.

\begin{figure}[t]
\centering
\hspace*{-.4cm}\includegraphics[width=0.55\textwidth,height=3in]{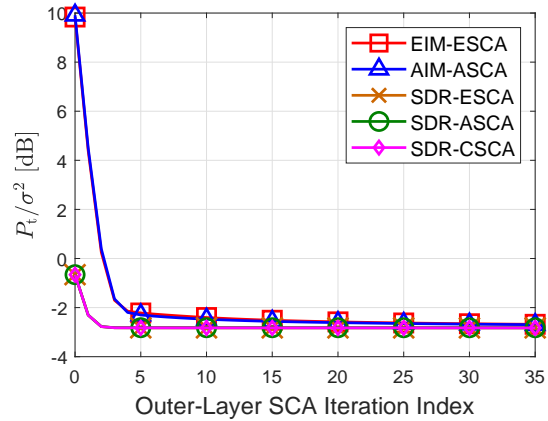}\vspace{-5em}
\caption{Convergence behavior for $\Pc_{o}$: Normalized power objective $P_{\rm t}/\sigma^2$ over the outer-layer SCA iterations ($N = 500$, $K = 10$).}\vspace*{-1em}
\label{Fig2:QoS_Convergence_Outer_Power}
\end{figure}

We now study the convergence behavior of different algorithms
over the outer-layer SCA iterations.
Fig.~\ref{Fig2:QoS_Convergence_Outer_Power} shows
the convergence behavior of our proposed algorithms
at the outer-layer SCA iterations, for $N = 500$ and $K = 10$.
We set the inner-layer convergence threshold for the proposed algorithms such that they converge to nearly the same value. For  EIM-ESCA and SDR-ESCA, we set the inner-layer convergence threshold to be
$\frac{\|\xbf^{(n+1)}-\xbf^{(n)}\|}{\|\xbf^{(n+1)}\|}$ $\leq 10^{-3}$.
Note that although  Fig.~\ref{Fig1:QoS_Convergence_Inner_Power} shows that at the first outer-layer SCA iteration,  EIM-ESCA converges to a higher value of $P_{\rm t}/\sigma^2$
than   that of SDR-ESCA over the inner-layer iterations,
Fig.~\ref{Fig2:QoS_Convergence_Outer_Power} shows that
EIM-ESCA  eventually converges to nearly the same value of $P_{\rm t}/\sigma^2$ as that of SDR-ESCA over the outer-layer SCA iterations.
For ASCA, we set the inner-layer convergence threshold
$\frac{\|\abf^{(n+1)}-\abf^{(n)}\|}{\|\abf^{(n+1)}\|}$ $\leq 10^{-3}$ for SDR-ASCA and  $\frac{\|\abf^{(n+1)}-\abf^{(n)}\|}{\|\abf^{(n+1)}\|}$ $\leq 0.2\times 10^{-3}$ for AIM-ASCA. Our experiments show that a tighter inner-layer threshold for AIM-ASCA than that of SDR-ASCA is needed  to converge to the same  value of $P_{\rm t}/\sigma^2$  as the rest of algorithms at the outer-layer iterations.
As we see in Fig.~\ref{Fig2:QoS_Convergence_Outer_Power},
all algorithms converge to the same value of the normalized transmit power $P_{\rm t}/\sigma^2$ within 35
SCA iterations.
Again, for different initialization methods,
 since the SDR method provides a better initial point than EIM  and AIM methods, it leads to a faster convergence for both ESCA and ASCA, where only less than 5 SCA iterations are required to reach convergence.
When the same SDR initialization is used, ESCA, ASCA, and CSCA  have a similar convergence behavior.

\subsection{Performance Comparison for the QoS Problem}

 \begin{figure}[t]
\centering
\hspace*{-.4cm}\includegraphics[width=0.55\textwidth,height=3in]{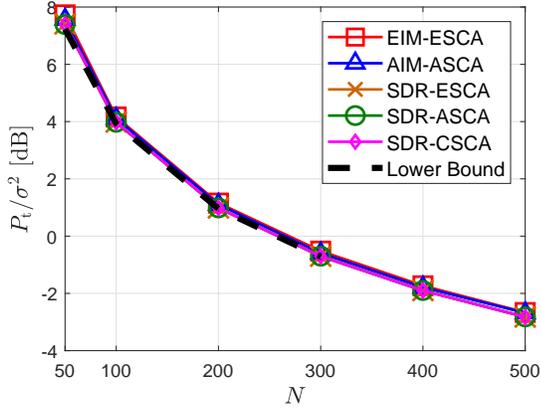}\vspace{-5em}
\caption{QoS: Normalized transmit power $P_{\rm t}/\sigma^{2}$ vs. $N$ ($G = 3$, $K = 10$).}  \vspace*{-1em}
\label{Fig4:QoS_Power_N}
\end{figure}

\begin{table}[t]
\footnotesize
\renewcommand{\arraystretch}{0}
\caption{QoS: Average Computation Time over $N$ (sec.) ($G = 3$, $K = 10$).}
\label{Table1:QoS_Compt_time_N}
\centering
\begin{tabular}{l| ccccccc}
\toprule
\hspace{3em}$N$                    &   100  &  200  &  300  & 400   &  500             \\ \midrule\midrule
EIM-ESCA                           &  2.694 & 3.032 & 2.708 & 2.589 & 2.711    \\ \midrule
AIM-ASCA                           &  1.150 & 2.415 & 3.195 & 4.092 & 4.339   \\ \midrule
SDR-ESCA                           &  0.327 & 0.237 & 0.233 & 0.224 & 0.243   \\ \midrule
SDR-ASCA                           &  0.076 & 0.051 & 0.061 & 0.071 & 0.088        \\ \midrule
SDR-CSCA \cite{Dong&Wang:TSP2020}  &  7.179 & 6.126 & 6.400 & 5.963 & 6.500        \\ \midrule\midrule
EIM (Init. method)                      &  0.037 & 0.044 & 0.045 & 0.050 & 0.057    \\ \midrule
AIM (Init. method)                      &  0.0068 & 0.0058 & 0.0059 & 0.0063 & 0.0073     \\ \midrule
SDR (Init. method)                      &  1.050 & 1.038 & 1.104  &  1.103  &  1.137      \\ \bottomrule
\end{tabular}\vspace*{-1em}
\end{table}

\begin{figure}[t]
\centering
\hspace*{-.4cm}\includegraphics[width=0.55\textwidth,height=3in]{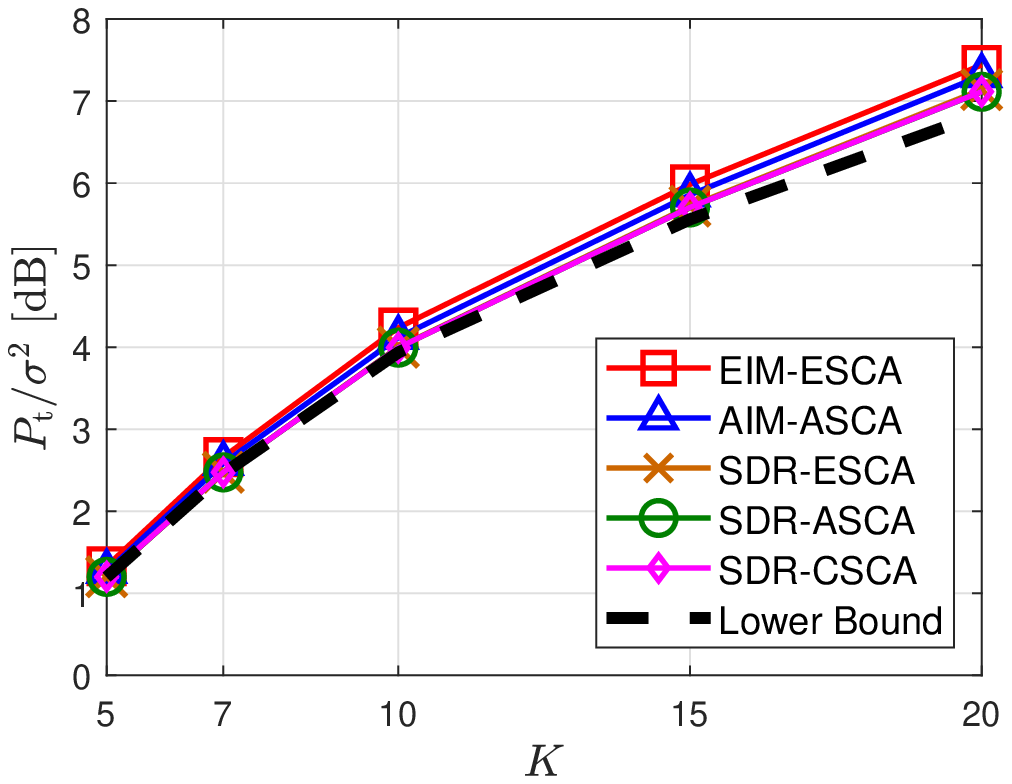}\vspace{-5em}
\caption{QoS: Normalized transmit power $P_{\rm t}/\sigma^{2}$ vs. $K$ ($G = 3$, $N = 100$).} \vspace*{-.5em}
\label{Fig6:QoS_Power_K}
\end{figure}
\begin{table}[t]
\scriptsize
%\footnotesize
\renewcommand{\arraystretch}{0}
\centering
\caption{QoS: Average Computation Time over $K$ (sec.) ($G = 3$, $N = 100$).}
\begin{tabular}{l| cccccccc}
\toprule
\hspace{3em}$K$                    &   5  &  7  &  10  & 15   &  20    & {35}                   \\ \midrule\midrule
EIM-ESCA                           & 1.098  &  1.689   & 2.737  &  5.498  &  8.685  &  {14.33 }    \\ \midrule
AIM-ASCA                           &  0.519  &   0.725   &  1.234 &  2.043  &  3.146  &  {5.426}    \\ \midrule
SDR-ESCA                           & 0.056  &    0.112  &    0.373  &   0.856  &  1.666  & {4.883}  \\ \midrule
SDR-ASCA                           & 0.012  &   0.024  &  0.087  &  0.216  &  0.433  &  {2.594}    \\ \midrule
SDR-CSCA \cite{Dong&Wang:TSP2020}  & 1.747  &  3.502   &   7.592 &  13.79 &   21.52  &  {57.85} \\ \midrule\midrule
EIM (Init. method)                      &  0.019  &  0.025  &  0.038  &  0.063  &  0.090    &  {0.450}  \\ \midrule
AIM (Init. method)                      & 0.0048  &  0.0055 &   0.0067  &  0.012  &    0.018  & {0.077}  \\ \midrule
SDR (Init. method)                      & 0.545   &  0.727  &    1.094  &  1.852  &  3.489    & {22.79}  \\ \bottomrule
\end{tabular}
\label{Table2:QoS_Compt_time_K}\vspace*{-.5em}
\end{table}

We now compare the performance of different algorithms
 for the QoS problem.
We set SINR target $\gamma = 10~\text{dB}$.
Also, for all algorithms in comparison,
we set the convergence threshold for the outer-layer SCA iterations as
$\frac{\|\vbf^{(l+1)}-\vbf^{(l)}\|}{\|\vbf^{(l+1)}\|}$ $\leq 10^{-3}$.
Fig.~\ref{Fig4:QoS_Power_N} shows the normalized transmit power $P_{\rm t}/\sigma^{2}$
 vs.\ $N$ by different algorithms, for $G = 3$ and $K = 10$.
 We see that our proposed algorithms have a similar performance to that of SDR-CSCA in \cite{Dong&Wang:TSP2020},
 and all algorithms  nearly attain the lower bound  for $\Pc_{o}$.\footnote{Note that  the lower bound is only shown until $N=300$ in Fig.~\ref{Fig4:QoS_Power_N} due to the high computational complexity involved in  generating the lower bound, which increases fast with $N$ and becomes impractical for
$N$ beyond 300. Similarly, the upper bound shown in Fig.~\ref{Fig7:MMF_SINR_N} is provided until $N=200$. }
 This demonstrates that our proposed fast algorithms achieve a near-optimal performance.
 The computational advantages of ESCA and ASCA are shown in Table~\ref{Table1:QoS_Compt_time_N},
 where we list the average computation time by each algorithm used for the plots in Fig.~\ref{Fig4:QoS_Power_N}.
The first five rows show the computation times of different algorithms,
excluding the initialization.
  We observe that, by using the optimal  structure in \eqref{optimal_QoS_structure},
 the computation times of all algorithms remain roughly unchanged as $N$ increases.
Under the same SDR initialization method, the computation times of SDR-ESCA and SDR-ASCA are
 only about $4 \%$ and $1 \%$ of that of SDR-CSCA, respectively,
 and SDR-ASCA has a smaller computation time than SDR-ESCA.
 The computation times of EIM-ESCA and AIM-ASCA  are both about $40\%$ of that of SDR-CSCA.
 The computation time of AIM-ASCA increases with $N$ more noticeably than other algorithms.
 As a result, AIM-ASCA is initially faster than EIM-ESCA for $N \le 200$,
 but its computation time increases and becomes slower than EIM-ESCA for $N\geq 300$. The reason is due to the quality of initialization as will be explained below.
The last three rows in Table~\ref{Table1:QoS_Compt_time_N}
show
 the computation times of different initialization methods.
We see that the EIM is a fast initialization method, with its computation time about $4 \%$ of that of SDR.
The AIM is the fastest one among all three initialization methods,
with its computation time about $15\%$ and $0.5\%$ of that of EIM and SDR, respectively.
 However, the quality of AIM initialization deteriorates as $N$ increases,
 unlike other initialization methods, leading to more iterations for convergence and a longer computation time. This is evidenced by the computation time of AIM-ASCA, which increases more noticeably as $N$ increases.

Fig.~\ref{Fig6:QoS_Power_K} shows $P_{\rm t}/\sigma^{2}$
 vs.\ $K$ by different algorithms, for $G = 3$ and $N = 100$.
 Again, the performance of SDR-ESCA and SDR-ASCA are nearly identical to that of SDR-CSCA
 and nearly attain the lower bound.
 The performance gaps of EIM-ESCA and AIM-ASCA to the lower bound are more noticeable as $K$ becomes large,
 although they are still within $0.6~\text{dB}$.
  Similar to Table~\ref{Table1:QoS_Compt_time_N},
 the corresponding average computation times of these algorithms are shown in Table~\ref{Table2:QoS_Compt_time_K}.
 We see that both ESCA and ASCA are considerably faster than CSCA to compute the solution.
 In particular, unlike CSCA, their computation times only mildly increase  with $K$.
For the initialization methods,
EIM and AIM  are faster  and   much more scalable over $K$
than SDR. The complexity of SDR increases noticeably over $K$, and thus, it is not  a suitable initialization method for very large systems.
Overall, in terms of the total computation
time (including initialization and the algorithm itself),
for $N=100$ and $K\leq 20$, SDR-ASCA is the fastest  one among all algorithms.
AIM-ASCA  offers comparable computation time as SDR-ASCA.  For $N=100$ and $K\ge 35$, AIM-ASCA and  EIM-ESCA offers faster computation time than the rest using SDR for initialization.

\emph{ Comparison Summary}: Based on the above simulation analysis, we have the following comparison summary  of the two proposed algorithms:
\begin{itemize}
\item  Among the initialization methods (EIM, AIM, and SDR), SDR\ provides the best initialization point, which leads to faster computation for both ESCA and ASCA (\ie SDR-ESCA and SDR-ASCA). However, the computational complexity of SDR is high. As  $K$ becomes large (\eg $K> 20$),
SDR-based initialization becomes computationally expensive and is not recommended. EIM and AIM are both very low-complexity methods over $N$ and $K$. AIM provides faster initialization than EIM.
However, the quality of the initial point that AIM provides deteriorates over $N$, leading to a noticeable increase of computation time by AIM-ASCA over $N$. In contrast, the computation time of EIM-ESCA remains roughly unchanged over $N$.
%%%%%%%%%%%%%%%%%%%%%%%
\item Both ESCA and ASCA provide closed-form updates in each iteration. The computational complexity of  ESCA and ASCA   grow over $K$.
From our study, we conclude that:
\begin{itemize}\item For the system with users per group $K\le15$, SDR-ASCA is generally
 the fastest among all algorithms. When $N \le 100$, AIM-ASCA  is similar to SDR-ASCA. However, the complexity of SDR-ASCA increases over $K,$ and that of AIM-ASCA increases over $N$. As both $K$ and $N$ grow, SDR-ASCA and AIM-ASCA have higher computation time and are no longer preferred.
 \item For a large-scale system where both $N$ and $K$ are large (\eg $N\ge500$, $K\ge 35$),
EIM-ESCA is expected to provide the
fastest computation time among these  algorithms and thus is preferred.
 \end{itemize}
  \end{itemize}

\subsection{Performance Comparison for the MMF Problem}

\begin{figure}[t]
\centering
\hspace*{-.4cm}\includegraphics[width=0.55\textwidth,height=3in]{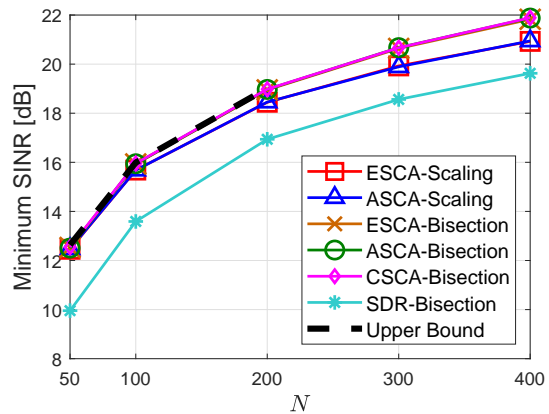}\vspace{-5em}
\caption{MMF: Minimum SINR vs. $N$ ($G = 3$, $K = 10$).} \vspace*{-.5em}
\label{Fig7:MMF_SINR_N}
\end{figure}
\begin{table}[t]
\footnotesize
\renewcommand{\arraystretch}{0}
\caption{MMF: Average Computation Time over $N$ (sec.) ($G = 3$, $K = 10$).}
\centering
\begin{tabular}{l| cccccccc}
\toprule
\hspace{3em}$N$                        &   50   &   100   &   200  &  300  & 400             \\ \midrule\midrule
ESCA-Scaling                            &  0.482 &  0.327 & 0.237 & 0.233  & 0.224         \\ \midrule
ASCA-Scaling                            &  0.099 &  0.076 & 0.051 & 0.061   & 0.071    \\ \midrule
ESCA-Bisection                           &  16.43 &  21.11 & 33.02 & 46.18  &  44.63          \\ \midrule
ASCA-Bisection                           &  12.35 &  12.11 & 13.06 & 14.09  &  17.94        \\ \midrule
CSCA-Bisection \cite{Dong&Wang:TSP2020}  &  99.04 &  87.33 & 81.62 & 84.10  &  99.86    \\ \midrule
SDR-Bisection      \cite{Dong&Wang:TSP2020}  &  11.19 &  15.46 & 19.85 & 15.82 &  22.12          \\ \bottomrule
\end{tabular}
\label{Table3:MMF_Compt_time_N}\vspace*{-.6em}
\end{table}

\begin{figure}[t]
\centering
\hspace*{-.4cm}\includegraphics[width=0.55\textwidth,height=3in]{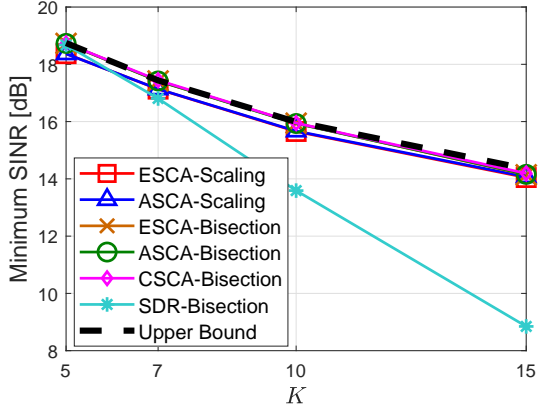}\vspace{-5em}
\caption{MMF: Minimum SINR vs. $K$ ($G = 3$, $N = 100$).}
\label{Fig8:MMF_SINR_K}
\end{figure}

We now present the performance
of our proposed Algorithm~\ref{alg:MMF}
 for the MMF problem $\Sc_o$.
 We set the maximum transmit power budget against noise variance as $P/\sigma^2=10~\text{dB}$.
Fig.~\ref{Fig7:MMF_SINR_N} plots the average minimum SINR vs. $N$,
and Table~\ref{Table3:MMF_Compt_time_N} shows the corresponding computation time by these algorithms,
for $G = 3$ and $K = 10$.
Both ESCA-Bisection and ASCA-Bisection provide near-identical performance
to that of CSCA-Bisection,
and they are nearly optimal as compared with the upper bound,
but with much lower computation times.
The proposed simple ESCA-Scaling and ASCA-Scaling algorithms for the MMF problem
nearly attain the upper bound for $N\leq 100$.
Their performance gap to the upper bound increases as $N$ increases,
indicating the accuracy of the scaling degrades as $N$ becomes large.
At $N = 400$, the gap is about $1~\text{dB}$.
Nonetheless,
the ESCA-Scaling and ASCA-Scaling are several orders of magnitude faster than
ESCA-Bisection and ASCA-Bisection.
SDR-Bisection has  worse performance than all the rest algorithms.
This is because,  for the QoS problem $\Pc_{1}$  at each bi-section iteration,
the SDR approximation deteriorates when the number of constraints ($GK$) for $\Pc_{1}$ is large.
Finally,
Fig.~\ref{Fig8:MMF_SINR_K} shows the average minimum SINR vs. $K$
by different algorithms,
for $G = 3$ and $N = 100$, with the corresponding computation time shown in  Table~\ref{Table4:MMF_Compt_time_K}.
Except for the SDR-Bisection, all the proposed algorithms nearly attain the upper bound
 and thus are nearly optimal.
 In particular, ESCA-Scaling and ASCA-Scaling maintain their near-optimal performance as $K$ increases.
The computational advantage of ESCA-Scaling and ASCA-Scaling is clearly seen in Table~\ref{Table4:MMF_Compt_time_K},
where both algorithms are substantially faster in computing a solution than the other algorithms.
\begin{table}[t]
\footnotesize
\renewcommand{\arraystretch}{0}
\centering
\caption{MMF: Average Computation Time over $K$ (sec.) ($G = 3$, $N = 100$).}
\begin{tabular}{l| ccccccc}
\toprule
\hspace{3em}$K$                        &   5    &   7    &  10   &  15              \\ \midrule\midrule
ESCA-Scaling                            &  0.056  &    0.112  &    0.373  &   0.856  \\ \midrule
ASCA-Scaling                            &  0.012  &   0.024  &  0.087  &  0.216      \\ \midrule
ESCA-Bisection                           &  7.297 & 12.62  & 21.26 &  39.59           \\ \midrule
ASCA-Bisection                           &  6.192 & 8.628  & 12.13 &  25.18           \\ \midrule
CSCA-Bisection \cite{Dong&Wang:TSP2020}  &  24.91 & 48.47  & 88.40 &  178.5           \\ \midrule
SDR-Bisection      \cite{Dong&Wang:TSP2020}  &  6.300 & 10.10  & 15.53 &  25.75           \\ \bottomrule
\end{tabular}
\label{Table4:MMF_Compt_time_K}\vspace*{-1em}
\end{table}

\section{Conclusion}
\label{sec:conclusion}
In this work, exploiting the optimal multicast beamforming structure,
we proposed  two fast computational algorithms,  ESCA and ASCA, for multi-group multicast beamforming design.
For the QoS problem solved by the SCA method, these two algorithms provide
efficient computational methods for solving the convex subproblems of SCA.
At each SCA iteration,
ESCA implements a dual saddle point reformulation along with the extragradient method
to solve the convex subproblem;
ASCA constructs an
ADMM procedure in a form that decomposes the convex subproblem into multiple smaller subproblems for parallel computing
with closed-form updates.
To provide effective initial feasible points for SCA to facilitate fast convergence,
we proposed three initialization methods based on the extragradient method,
ADMM, and SDR.
For the MMF problem,
we further proposed a simple closed-form scaling approach based on the solution to the QoS problem, avoiding the high computational complexity involved in iteratively solving the QoS problems,
while offering bounded performance guarantee.
Our simulation studies showed that the proposed ESCA and ASCA
provide near-optimal performance with substantially lower computational complexity than the state-of-the-art algorithms for large-scale systems.

Finally, note that in this work, we assumed single-antenna users in our system model  for multicast beamforming design. For the case of multi-antenna users,
with a given linear receiver processing technique  implemented by the receiver, each MIMO channel can be converted into an  equivalent MISO channel.
Therefore, our proposed  algorithms can be  relatively straightforward  to be applied to the case of multi-antenna users with their receiver processing techniques given.

\appendices
\section{Derivation of  $\dbf^{o}_{ik}$ in \eqref{eq_d_optimal}}
\label{append:deriv_closed_d}

The Lagrangian of $\Pc_{\text{dsub}}({\bf{v}})$ is given by
\vspace*{-.5em}
\begin{align}
\Lc(\dbf_{ik}, \nu_{ik}) = &  \sum_{j=1}^{G} |d_{jik}-e^{(n)}_{1,jik}|^{2} +  \nu_{ik}e_{2,ik} \nn\\
& + \nu_{ik}\gamma_{ik}\sum_{j\neq i}|d_{jik}|^{2}-2\nu_{ik}\Re{\{d_{iik}e_{3,ik}\}} \nn
\end{align}
where $\nu_{ik} \geq 0$ is the Lagrange multiplier associated with constraint \eqref{d_constraint_2}.
Since $\Pc_{\text{dsub}}({\bf{v}})$ is convex,
the optimal  $\dbf^{o}_{ik}$ and $\nu^{o}_{ik}$ satisfy the KKT conditions \cite{Boyd&Vandenberghe:book2004}.
Setting the gradient
$\nabla_{\dbf_{ik}}\Lc(\dbf_{ik}, \nu^{o}_{ik}) = 0$,
we obtain $\dbf^{o}_{ik}$  in \eqref{eq_d_optimal}.
Substituting the expression of  $d_{jik}^o$ in \eqref{eq_d_optimal} into the constraint of $\Pc_{\text{dsub}}({\bf{v}})$ yields
\vspace*{-1em}
\begin{align}
& f(\nu^{o}_{ik})\triangleq e_{2,ik}+\gamma_{ik}\sum_{j\neq i}\frac{|e^{(n)}_{1,jik}|^{2}}{(1+\nu^{o}_{ik}\gamma_{ik})^{2}}-2\Re{\{e_{3,ik}e^{(n)}_{1,iik}\}}\nn\\
& \qquad\qquad -2\nu^{o}_{ik}|e_{3,ik}|^{2} \leq 0.\nn
\end{align}
It is shown in \cite[Appendix A]{Chen&Tao:ITC2017} that
the function $f(\nu^{o}_{ik})$ is strictly decreasing for $\nu^{o}_{ik}\geq 0$.
By the complementary slackness condition, we have
$\nu^{o}_{ik}f(\nu^{o}_{ik}) = 0$.
If $f(0)\leq 0$, then $f(\nu^{o}_{ik})<0$ for any $\nu^{o}_{ik}\geq 0$,
 and   we have $\nu^{o}_{ik} = 0$.
Otherwise,
$\nu^{o}_{ik} = 0$ is not an feasible dual solution for $\Pc_{\text{dsub}}({\bf{v}})$; thus,  $f(\nu^{o}_{ik}) = 0$,
and $\nu^{o}_{ik}$ is the unique real positive root of the cubic equation
\eqref{eq_cubic},
whose roots can be obtained  by the cubic formula \cite{Chen&Tao:ITC2017}.

\section{Closed-Form Updating Steps for AIM}
\label{append:derivation_AIM}

\subsubsection{Closed-Form $\dbf$-update in \eqref{d_update_AIM}}
\vspace*{-.4em}

Similar to that in Section~\ref{sec_ADMM_d_update},
given $\abf^{(n)}$ and $\tqbf^{(n)}$,
the optimization problem in \eqref{d_update_AIM} can be decomposed into $K_{\text{tot}}$ subproblems,
one for each user $k$ in group $i$ given by
\vspace*{-.5em}
\begin{align}
\Pc_{\text{dsub}}^{\prime}: & \min_{\dbf_{ik}}\,\,\sum_{j=1}^{G}|d_{jik}-\te^{(n)}_{1,jik}|^{2}  \nn\\
               & \,\, {\rm s.t.}\,\,\gamma_{ik}\sum_{j\neq i}|d_{jik}|^{2}+\gamma_{ik}\sigma^2-|d_{iik}|^{2}\leq 0.
\label{AIM_subproblem}
\end{align}
where $\te^{(n)}_{1,jik}\triangleq\abf^{(n)H}_{j}\fbf_{jik} - \tq^{(n)}_{jik}$,
for $k\in\Kc_{i},i,j\in\Gc$.
Problem $\Pc_{\text{dsub}}^{\prime}$ is a non-convex QCQP-1 problem
similar to $\Pc_{\text{dsub}}({\bf{v}})$.
Since it satisfies Slater's condition, the strong duality holds \cite[Appendix B]{Boyd&Vandenberghe:book2004}, and
the optimal $\dbf_{ik}^o$ satisfies the KKT conditions.
Let $\mu_{ik}^o\ge0$ be the optimal Lagrange multiplier associated with constraint \eqref{AIM_subproblem}. For  $\Pc_{\text{dsub}}^{\prime}$ being feasible, we have  $0\le\mu_{ik}^o\le1$ \cite[Appendix B]{Boyd&Vandenberghe:book2004}.  Thus, using the KKT conditions and with a procedure similar to that  in Appendix~\ref{append:deriv_closed_d},
we have
the closed-form optimal solution $\dbf_{ik}$ to $\Pc_{\text{dsub}}^{\prime}$
given by
\begin{align}
d^{o}_{jik} =
\begin{cases}
\frac{\te^{(n)}_{1,iik}}{1-\mu^{o}_{ik}}           & \text{if} \,\,\, j = i\text{,} \nn\\
\frac{\te^{(n)}_{1,jik}}{1+\mu^{o}_{ik}\gamma_{ik}}           & \text{otherwise}\text{}\nn
\end{cases}
\end{align}
where   $\mu^{o}_{ik} = 0$ if $\gamma_{ik}\sum_{j\neq i}|\te^{(n)}_{1,jik}|^{2}+\gamma_{ik}\sigma^2-|\te^{(n)}_{1,iik}|^{2}\leq0$;
otherwise the following quartic equation is guaranteed to have a unique root in $(0,1)$, and $\mu^{o}_{ik}$ is this unique root:
\begin{align}
\gamma_{ik}\frac{\sum_{j\neq i}|\te^{(n)}_{1,jik}|^{2}}{(1+\mu^{o}_{ik}\gamma_{ik})^{2}}+\gamma_{ik}\sigma^2-\frac{|\te^{(n)}_{1,iik}|^{2}}{(1-\mu^{o}_{ik})^{2}} = 0.\nn
\end{align}

\subsubsection{Closed-Form $\abf$-update in \eqref{a_update_AIM}}
Given $\dbf^{(n+1)}$ and $\tqbf^{(n)}$,
the optimization problem in \eqref{a_update_AIM}
can be decomposed into $G$ subproblems,
one for each group $j$ given by
\begin{align}
\Pc^{\prime}_{\text{asub}}: \min_{\abf} \sum_{i=1}^{G}\sum_{k=1}^{K_i}|d^{(n+1)}_{jik}\!-\!\abf^{H}_{j}\fbf_{jik}\!+\!\tq^{(n)}_{jik}|^{2},\nn
\end{align}
which  is an unconstrained quadratic convex optimization problem.
The closed-form solution to $\Pc^{\prime}_{\text{asub}}$ is expressed as
\begin{align}
\abf^{(n+1)}_{j}\!=\! \left(\sum_{i=1}^{G}\sum_{k=1}^{K_i}\fbf_{jik}\fbf^{H}_{jik}\right)^{-1}\!\sum_{i=1}^{G}\sum^{K_i}_{k=1}\left(d^{(n+1)}_{jik}\!+\!\tq^{(n)}_{jik}\right)^{*}\fbf_{jik}.\nn
\end{align}

\section{Proof of Proposition~\ref{thm:duality}}
\label{append:proof_thm_1}

\vspace*{-.5em}

\begin{IEEEproof}
It is straightforward to check that the scaled beamforming vector $\wbf^{\text{s}}(\gammabf, P)=$ $\sqrt{\frac{P}{P^{\sssQ}(\gammabf)}}\wbf^{\sssQ}(\gammabf)$
satisfies constraint \eqref{MMF_constraint}
and therefore is a feasible solution to $\mathcal{S}_o(\gammabf, P)$.
The achieved objective value $t^{\text{s}}(\gammabf, P)$ corresponding to
$\wbf^{\text{s}}(\gammabf, P)$
is expressed in terms of $\wbf^{\sssQ}(\gammabf)$ as
\begin{align}
t^{\text{s}}(\gammabf,P) & = \min_{i,k} \,  \frac{1}{\gamma_{ik}}\frac{\frac{P}{P^{\sssQ}(\gammabf)}|\hbf_{ik}^{H}\wbf_i^{\sssQ}(\gammabf)|^{2}}{\frac{P}{P^{\sssQ}(\gammabf)}I^{\sssQ}_{ik}(\gammabf)+\sigma^{2}}\nonumber\\
& = \min_{i,k} \,  \frac{\frac{P}{P^{\sssQ}(\gammabf)}|\hbf_{ik}^{H}\wbf_i^{\sssQ}(\gammabf)|^{2}}{{\gamma_{ik}I^{\sssQ}_{ik}(\gammabf)+\gamma_{ik}\sigma^2}} \frac{{I^{\sssQ}_{ik}(\gammabf)+\sigma^2}}{\frac{P}{P^{\sssQ}(\gammabf)}I^{\sssQ}_{ik}(\gammabf)+\sigma^{2}}.
\label{t_express}
\end{align}
Define
$t^{\sssQ}(\gammabf)$ $\triangleq$ $\min_{i,k} \, \frac{|\hbf_{ik}^{H}\wbf_i^{\sssQ}(\gammabf)|^{2}}{\gamma_{ik}I^{\sssQ}_{ik}(\gammabf)+\gamma_{ik}\sigma^2}$.
Since $\wbf^{\sssQ}(\gammabf)$ satisfies constraint \eqref{SINR_constraint} in $\mathcal{P}_{o}(\gammabf)$
as a feasible solution,
we have $\frac{|\hbf_{ik}^{H}\wbf_i^{\sssQ}(\gammabf)|^{2}}{I^{\sssQ}_{ik}(\gammabf)+\sigma^2}$
$\geq$ $\gamma_{ik}$ for $k\in \Kc_{i}, i\in \Gc$.
It follows that $t^{\sssQ}(\gammabf)\geq 1$.
Based on this,
from \eqref{t_express},
we have
\begin{align}
t^{\text{s}}(\gammabf, P) & \geq \frac{P}{P^{\sssQ}(\gammabf)}t^{\sssQ}(\gammabf) \min_{i,k} \, \frac{{I^{\sssQ}_{ik}(\gammabf)+\sigma^2}}{\frac{P}{P^{\sssQ}(\gammabf)}I^{\sssQ}_{ik}(\gammabf)+\sigma^{2}} \nn\\
&  \geq  \frac{P}{P^{\sssQ}(\gammabf)}\min_{i,k} \, \frac{{I^{\sssQ}_{ik}(\gammabf)+\sigma^2}}{\frac{P}{P^{\sssQ}(\gammabf)}I^{\sssQ}_{ik}(\gammabf)+\sigma^{2}}.
\label{ineq_lower_bound}
\end{align}
From \eqref{t_express},
we can also obtain an upper bound on $t^{\text{s}}(\gammabf, P)$ as
\begin{align}
t^{\text{s}}(\gammabf, P) & \leq
\frac{P}{P^{\sssQ}(\gammabf)}t^{\sssQ}(\gammabf) \frac{{I^{\sssQ}_{i'\!k'}(\gammabf)+\sigma^2}}{\frac{P}{P^{\sssQ}(\gammabf)}I^{\sssQ}_{i'\!k'}(\gammabf)+\sigma^{2}} \nn\\
& \leq
\frac{P}{P^{\sssQ}(\gammabf)}t^{\sssQ}(\gammabf) \max_{i,k} \, \frac{{I^{\sssQ}_{ik}(\gammabf)+\sigma^2}}{\frac{P}{P^{\sssQ}(\gammabf)}I^{\sssQ}_{ik}(\gammabf)+\sigma^{2}}
\label{ineq_upper_bound}
\end{align}
where $\{i',k'\}$ $=$ $\underset{i,k}{\arg\min}\, \frac{|\hbf_{ik}^{H}\wbf_i^{\sssQ}(\gammabf)|^{2}}{\gamma_{ik}I^{\sssQ}_{ik}(\gammabf)+\gamma_{ik}\sigma^2}$.
Note that the scaled beamforming vector
$\frac{\wbf^{\sssQ}(\gammabf)}{\sqrt{t^{\sssQ}(\gammabf)}}$
results in the transmit power $\frac{P^{\sssQ}(\gammabf)}{t^{\sssQ}(\gammabf)}$.
Then, with $\frac{\wbf^{\sssQ}(\gammabf)}{\sqrt{t^{\sssQ}(\gammabf)}}$,
the achieved weighted SINR,
for any $k\in \mathcal{K}_i, i\in \Gc$, is given by
\begin{align} \frac{1}{\gamma_{ik}}\frac{\frac{1}{t^{\sssQ}(\gammabf)}|\hbf_{ik}^{H}\wbf_i^{\sssQ}(\gammabf)|^{2}}{{\frac{1}{t^{\sssQ}(\gammabf)}I^{\sssQ}_{ik}(\gammabf)+\sigma^2}} \stackrel{(a)}{\geq}
\frac{1}{\gamma_{ik}t^{\sssQ}(\gammabf)}\frac{|\hbf_{ik}^{H}\wbf_i^{\sssQ}(\gammabf)|^{2}}{{I^{\sssQ}_{ik}(\gammabf)+\sigma^2}}
\stackrel{(b)}{\geq} 1
\label{ineq_scaling}
\end{align}
where $(a)$ is due to $t^{\sssQ}(\gammabf)\geq 1$,
and $(b)$ is from the definition of $t^{\sssQ}(\gammabf)$.
Thus,
the beamforming vector $\frac{\wbf^{\sssQ}(\gammabf)}{\sqrt{t^{\sssQ}(\gammabf)}}$
is feasible to $\mathcal{P}_{o}(\gammabf)$
and
$\frac{P^{\sssQ}(\gammabf)}{t^{\sssQ}(\gammabf)}$ $\geq$ $P^{o}(\gammabf)$,
where $P^{o}(\gammabf)$ is the minimum transmit power in $\mathcal{P}_{o}(\gammabf)$.
Applying $\frac{P^{\sssQ}(\gammabf)}{t^{\sssQ}(\gammabf)}$ $\geq$ $P^{o}(\gammabf)$
to the RHS of \eqref{ineq_upper_bound} yields
\vspace*{-.5em}
\begin{align}
t^{\text{s}}(\gammabf, P) \leq \frac{P}{P^{o}(\gammabf)} \max_{i,k} \, \frac{{I^{\sssQ}_{ik}(\gammabf)+\sigma^2}}{\frac{P}{P^{\sssQ}(\gammabf)}I^{\sssQ}_{ik}(\gammabf)+\sigma^{2}}.
\label{ineq_upper_bound_final}
\end{align}
Combining \eqref{ineq_lower_bound} and \eqref{ineq_upper_bound_final},\
we have \eqref{scaling_bound}.
\end{IEEEproof}

\balance


\begin{thebibliography}{10}
\providecommand{\url}[1]{#1}
\csname url@samestyle\endcsname
\providecommand{\newblock}{\relax}
\providecommand{\bibinfo}[2]{#2}
\providecommand{\BIBentrySTDinterwordspacing}{\spaceskip=0pt\relax}
\providecommand{\BIBentryALTinterwordstretchfactor}{4}
\providecommand{\BIBentryALTinterwordspacing}{\spaceskip=\fontdimen2\font plus
\BIBentryALTinterwordstretchfactor\fontdimen3\font minus
  \fontdimen4\font\relax}
\providecommand{\BIBforeignlanguage}[2]{{%
\expandafter\ifx\csname l@#1\endcsname\relax
\typeout{** WARNING: IEEEtran.bst: No hyphenation pattern has been}%
\typeout{** loaded for the language `#1'. Using the pattern for}%
\typeout{** the default language instead.}%
\else
\language=\csname l@#1\endcsname
\fi
#2}}
\providecommand{\BIBdecl}{\relax}
\BIBdecl

\bibitem{Zhang&etal:2021ICASSP}
C.~{Zhang}, M.~{Dong}, and B.~{Liang},
\newblock ``First-order fast algorithm for structurally optimal multi-group
  multicast beamforming in large-scale systems,''
\newblock in {\em Proc. IEEE Int. Conf. Acoust., Speech, Signal Process.}, Jun.
  2021, pp. 4790--4794.

\bibitem{Dong&Wang:TSP2020}
M.~{Dong} and Q.~{Wang},
\newblock ``Multi-group multicast beamforming: Optimal structure and efficient
  algorithms,''
\newblock {\em {IEEE} Trans. Signal Process.}, vol. 68, pp. 3738--3753, May
  2020.

\bibitem{Larsson&Edfors&Tufvesson&Marzetta:ICM:14}
E.~G. Larsson, O.~Edfors, F.~Tufvesson, and T.~L. Marzetta,
\newblock ``Massive {MIMO} for next generation wireless systems,''
\newblock {\em {IEEE} Commun. Mag.}, vol. 52, no. 2, pp. 186--195, Feb. 2014.

\bibitem{Sidiropoulos&etal:TSP2006}
N.~D. {Sidiropoulos}, T.~N. {Davidson}, and Z.-Q. {Luo},
\newblock ``Transmit beamforming for physical-layer multicasting,''
\newblock {\em {IEEE} Trans. Signal Process.}, vol. 54, no. 6, pp. 2239--2251,
  Jun. 2006.

\bibitem{Karipidis&etal:TSP2008}
E.~{Karipidis}, N.~D. {Sidiropoulos}, and Z.-Q. {Luo},
\newblock ``Quality of service and max-min fair transmit beamforming to
  multiple cochannel multicast groups,''
\newblock {\em {IEEE} Trans. Signal Process.}, vol. 56, no. 3, pp. 1268--1279,
  Mar. 2008.

\bibitem{Ottersten&etal:TSP14}
D.~Christopoulos, S.~Chatzinotas, and B.~Ottersten,
\newblock ``Weighted fair multicast multigroup beamforming under per-antenna
  power constraints,''
\newblock {\em {IEEE} Trans. Signal Process.}, vol. 62, no. 19, pp. 5132--5142,
  Oct. 2014.

\bibitem{Jordan&Gong&Ascheid:Globecom09}
M.~Jordan, X.~Gong, and G.~Ascheid,
\newblock ``Multicell multicast beamforming with delayed {SNR} feedback,''
\newblock in {\em 2009 IEEE Global Telecommun. Conf.}, Nov. 2009, pp. 1--6.

\bibitem{Xiang&Tao&Wang:IJWC:13}
Z.~Xiang, M.~Tao, and X.~Wang,
\newblock ``Coordinated multicast beamforming in multicell networks,''
\newblock {\em {IEEE} Trans. Wireless Commun.}, vol. 12, no. 1, pp. 12--21,
  Jan. 2013.

\bibitem{Bornhorst&etal:2012TSP}
N.~{Bornhorst}, M.~{Pesavento}, and A.~B. {Gershman},
\newblock ``Distributed beamforming for multi-group multicasting relay
  networks,''
\newblock {\em IEEE Trans. Signal Process.}, vol. 60, no. 1, pp. 221--232, Jan.
  2012.

\bibitem{DongLiang:CAMSAP13}
M.~Dong and B.~Liang,
\newblock ``Multicast relay beamforming through dual approach,''
\newblock in {\em Proc. IEEE Int. Workshop Comput. Advances Multi-Sensor
  Adaptive Process.}, Dec. 2013, pp. 492--495.

\bibitem{Phan&etal:2009TSP}
K.~T. {Phan}, S.~A. {Vorobyov}, N.~D. {Sidiropoulos}, and C.~{Tellambura},
\newblock ``Spectrum sharing in wireless networks via {QoS}-aware secondary
  multicast beamforming,''
\newblock {\em IEEE Trans. Signal Process.}, vol. 57, no. 6, pp. 2323--2335,
  Jun. 2009.

\bibitem{Huang&etal:2012TSP}
Y.~{Huang}, Q.~{Li}, W.-K. {Ma}, and S.~{Zhang},
\newblock ``Robust multicast beamforming for spectrum sharing-based cognitive
  radios,''
\newblock {\em IEEE Trans. Signal Process.}, vol. 60, no. 1, pp. 527--533, Jan.
  2012.

\bibitem{Luo&etal:2010SPM}
Z.-Q. {Luo}, W.-K. {Ma}, A.~M.-C. {So}, Y.~{Ye}, and S.~{Zhang},
\newblock ``Semidefinite relaxation of quadratic optimization problems,''
\newblock {\em IEEE Signal Process. Mag.}, vol. 27, no. 3, pp. 20--34, May
  2010.

\bibitem{Marks&Wright:OperReas1978}
B.~R. {Marks} and G.~P. {Wright},
\newblock ``A general inner approximation algorithm for nonconvex mathematical
  programs,''
\newblock {\em Oper. Res.}, vol. 26, no. 4, pp. 681--683, Aug. 1978.

\bibitem{Tran&etal:SPL2014}
L.~{Tran}, M.~F. {Hanif}, and M.~{Juntti},
\newblock ``A conic quadratic programming approach to physical layer
  multicasting for large-scale antenna arrays,''
\newblock {\em {IEEE} Signal Process. Lett.}, vol. 21, no. 1, pp. 114--117,
  Jan. 2014.

\bibitem{Mehanna&etal:2015}
O.~{Mehanna}, K.~{Huang}, B.~{Gopalakrishnan}, A.~{Konar}, and N.~D.
  {Sidiropoulos},
\newblock ``Feasible point pursuit and successive approximation of non-convex
  {QCQP}s,''
\newblock {\em {IEEE} Signal Process. Lett.}, vol. 22, no. 7, pp. 804--808,
  Jul. 2015.

\bibitem{Christopoulos&etal:SPAWC15}
D.~Christopoulos, S.~Chatzinotas, and B.~Ottersten,
\newblock ``Multicast multigroup beamforming for per-antenna power constrained
  large-scale arrays,''
\newblock in {\em Proc. IEEE Int. Workshop Signal Process. Advances Wireless
  Commun.}, Jun. 2015, pp. 271--275.

\bibitem{Boyd&Vandenberghe:book2004}
S.~{Boyd} and L.~{Vandenberghe},
\newblock {\em Convex Optimization},
\newblock Cambridge, U.K.: Cambridge Univ. Press, 2004.

\bibitem{Konar&Sidiropoulos:TSP2017}
A.~{Konar} and N.~D. {Sidiropoulos},
\newblock ``Fast approximation algorithms for a class of non-convex {QCQP}
  problems using first-order methods,''
\newblock {\em {IEEE} Trans. Signal Process.}, vol. 65, no. 13, pp. 3494--3509,
  Jul. 2017.

\bibitem{Ibrahim&etal:TSP2020}
M.~S. {Ibrahim}, A.~{Konar}, and N.~D. {Sidiropoulos},
\newblock ``Fast algorithms for joint multicast beamforming and antenna
  selection in massive {MIMO},''
\newblock {\em {IEEE} Trans. Signal Process.}, vol. 68, pp. 1897--1909, Mar.
  2020.

\bibitem{Chen&Tao:ITC2017}
E.~{Chen} and M.~{Tao},
\newblock ``A{DMM}-based fast algorithm for multi-group multicast beamforming
  in large-scale wireless systems,''
\newblock {\em {IEEE} Trans. Commun.}, vol. 65, no. 6, pp. 2685--2698, Jun.
  2017.

\bibitem{Sadeghi&etal:TWC17}
M.~Sadeghi, L.~Sanguinetti, R.~Couillet, and C.~Yuen,
\newblock ``Reducing the computational complexity of multicasting in
  large-scale antenna systems,''
\newblock {\em {IEEE} Trans. Wireless Commun.}, vol. 16, no. 5, pp. 2963--2975,
  May 2017.

\bibitem{Yu&Dong:ICASSP18}
J.~Yu and M.~Dong,
\newblock ``Low-complexity weighted {MRT} multicast beamforming in massive
  {MIMO} cellular networks,''
\newblock in {\em Proc. IEEE Int. Conf. Acoust., Speech, Signal Process.,},
  Apr. 2018, pp. 3849--3853.

\bibitem{Yu&Dong:SPAWC18}
J.~Yu and M.~Dong,
\newblock ``Distributed low-complexity multi-cell coordinated multicast
  beamforming with large-scale antennas,''
\newblock in {\em Proc. IEEE Int. Workshop Signal Process. Advances Wireless
  Commun.}, Jun. 2018, pp. 1--5.

\bibitem{Xiang&etal:2014JSAC}
Z.~{Xiang}, M.~{Tao}, and X.~{Wang},
\newblock ``Massive {MIMO} multicasting in noncooperative cellular networks,''
\newblock {\em {IEEE} J. Sel. Areas Commun.}, vol. 32, no. 6, pp. 1180--1193,
  Jun. 2014.

\bibitem{Sadeghi&Yuen:2015Globecom}
M.~{Sadeghi} and C.~{Yuen},
\newblock ``Multi-cell multi-group massive {MIMO} multicasting: {A}n asymptotic
  analysis,''
\newblock in {\em Proc. IEEE Global Commun. Conf.}, Dec. 2015, pp. 1--6.

\bibitem{Tervo:TSP18}
O.~{Tervo}, L.~{Tran}, H.~{Pennanen}, S.~{Chatzinotas}, B.~{Ottersten}, and
  M.~{Juntti},
\newblock ``Energy-efficient multicell multigroup multicasting with joint
  beamforming and antenna selection,''
\newblock {\em {IEEE} Trans. Signal Process.}, vol. 66, no. 18, pp. 4904--4919,
  Sept. 2018.

\bibitem{Sadghi&etal:IEEE_J_WCOM18}
M.~{Sadeghi}, E.~{Bj\"{o}rnson}, E.~G. {Larsson}, C.~{Yuen}, and T.~{Marzetta},
\newblock ``Joint unicast and multi-group multicast transmission in massive
  {MIMO} systems,''
\newblock {\em {IEEE} Trans. Wireless Commun.}, vol. 17, no. 10, pp.
  6375--6388, Oct. 2018.

\bibitem{Mohammadietal:SPAWC21}
S.~{Mohammadi}, M.~{Dong}, and S.~{ShahbazPanahi},
\newblock ``Fast algorithm for joint unicast and multicast beamforming in
  large-scale systems,''
\newblock in {\em Proc. {IEEE} Workshop on Signal Processing Advances in
  Wireless Commun.(SPAWC)}, Sept. 2021, pp. 91--95.

\bibitem{Joudeh&Clerckx:2017TWC}
H.~{Joudeh} and B.~{Clerckx},
\newblock ``Rate-splitting for max-min fair multigroup multicast beamforming in
  overloaded systems,''
\newblock {\em IEEE Trans. Wireless Commun.}, vol. 16, no. 11, pp. 7276--7289,
  Nov. 2017.

\bibitem{Tervo&etal:2018SPAWC}
O.~{Tervo}, L.-N. {Tran}, S.~{Chatzinotas}, B.~{Ottersten}, and M.~{Juntti},
\newblock ``Multigroup multicast beamforming and antenna selection with
  rate-splitting in multicell systems,''
\newblock in {\em Proc. IEEE Int. Workshop Signal Process. Advances Wireless
  Commun.}, Jun. 2018, pp. 1--5.

\bibitem{Chen&etal:2020TVT}
H.~{Chen}, D.~{Mi}, B.~{Clerckx}, Z.~{Chu}, J.~{Shi}, and P.~{Xiao},
\newblock ``Joint power and subcarrier allocation optimization for multigroup
  multicast systems with rate splitting,''
\newblock {\em IEEE Trans. Veh. Technol.}, vol. 69, no. 2, pp. 2306--2310, Feb.
  2020.

\bibitem{Korpelevich:Matecon1976}
G.~M. {Korpelevich},
\newblock ``The extragradient method for finding saddle points and other
  problems,''
\newblock {\em Matecon}, vol. 13, pp. 35–49, 1977.

\bibitem{Facchinei&Pang:book2003}
F.~Facchinei and J.-S. Pang,
\newblock {\em Finite-Dimensional Variational Inequalities and Complementarity
  Problems},
\newblock New York, NY: Springer-Verlag, 2003.

\bibitem{Khobotov:USSR1987}
E.~N. {Khobotov},
\newblock ``Modification of the extra-gradient method for solving variational
  inequalities and certain optimization problems,''
\newblock {\em USSR Comput. Math. Math. Phys.}, vol. 27, no. 5, pp. 120--127,
  1987.

\bibitem{Marcotte:INFORM1991}
P.~{Marcotte},
\newblock ``Application of {K}hobotov's algorithm to varational inequalities
  and network equilibrium problems,''
\newblock {\em Inf. Syst. Oper. Res.}, vol. 29, no. 4, pp. 258--270, Nov. 1991.

\bibitem{Boyd&etal:book2011}
S.~Boyd, N.~Parikh, E.~Chu, B.~Peleato, and J.~Eckstein,
\newblock ``Distributed optimization and statistical learning via the
  alternating direction method of multipliers,''
\newblock {\em Found. Trends Mach. Learn.}, vol. 3, no. 1, pp. 1--122, 2011.
\end{thebibliography}
\end{document}